\documentclass[amsmath,amssymb,prl,superscriptaddress,twocolumn]{revtex4-2}

\usepackage[compact]{titlesec}
\usepackage[utf8]{inputenc}
\usepackage{graphicx}
\usepackage{dcolumn}
\usepackage{bm}
\usepackage{natbib}
\usepackage{float}
\usepackage{amsmath}
\usepackage{xcolor}

\makeatletter
\def\maketitle{
\@author@finish
\title@column\titleblock@produce
\suppressfloats[t]}
\makeatother

\newcommand{\beginsupplement}{
    \onecolumngrid
    \setcounter{page}{1}
    \setcounter{table}{0}
    \renewcommand{\thetable}{S\arabic{table}}
    \setcounter{figure}{0}
    \renewcommand{\thefigure}{S\arabic{figure}}
    \setcounter{equation}{0}
    \renewcommand{\theequation}{S\arabic{equation}}
    \setcounter{secnumdepth}{2}
}

\begin{document}

\title{Supercurrent through a single transverse mode in nanowire Josephson junctions}

\author{B. Zhang}
\thanks{These two authors contributed equally}
\affiliation{Department of Physics and Astronomy, University of Pittsburgh, Pittsburgh, PA, 15260, USA}
\affiliation{Department of Physics, The Pennsylvania State University, University Park, PA 16802, USA}

\author{Z. Li}
\thanks{These two authors contributed equally}
\affiliation{Department of Physics and Astronomy, University of Pittsburgh, Pittsburgh, PA, 15260, USA}

\author{H. Wu}
\affiliation{Department of Physics and Astronomy, University of Pittsburgh, Pittsburgh, PA, 15260, USA}

\author{M. Pendharkar}
\altaffiliation[Current address: ]{Department of Materials Science and Engineering, Stanford University, Stanford, CA, USA, 94305}
\affiliation{Electrical and Computer Engineering, University of California, Santa Barbara, CA, 93106, USA}

\author{C. Dempsey}
\affiliation{Electrical and Computer Engineering, University of California, Santa Barbara, CA, 93106, USA}

\author{J.S. Lee}
\altaffiliation[Current address: ]{Department of Physics and Astronomy, University of Tennessee, Knoxville, TN, 37996, USA}
\affiliation{California NanoSystems Institute, University of California Santa Barbara, Santa Barbara, CA, 93106, USA}

\author{S.D. Harrington}
\affiliation{Materials Department, University of California Santa Barbara, Santa Barbara, CA, 93106, USA}

\author{C.J. Palmstrøm}
\affiliation{Electrical and Computer Engineering, University of California, Santa Barbara, CA, 93106, USA}
\affiliation{California NanoSystems Institute, University of California Santa Barbara, Santa Barbara, CA, 93106, USA}
\affiliation{Materials Department, University of California Santa Barbara, Santa Barbara, CA, 93106, USA}

\author{S.M. Frolov}
\email{frolovsm@pitt.edu}
\affiliation{Department of Physics and Astronomy, University of Pittsburgh, Pittsburgh, PA, 15260, USA}

\begin{abstract}

Hybrid superconductor-semiconductor materials are fueling research in mesoscopic physics and quantum technology. Recently demonstrated smooth $\beta$-Sn superconductor shells, due to the increased induced gap, are expanding the available parameter space to new regimes. Fabricated on quasiballistic InSb nanowires, with careful control over the hybrid interface, Sn shells yield measurable switching currents even when nanowire resistance is of order 10k$\Omega$. In this regime Cooper pairs travel through a purely 1D quantum wire for at least part of their trajectory. Here, we focus on the evolution of proximity-induced supercurrent in magnetic field parallel to the nanowire. Long decay up to fields of 1T is observed. At the same time, the decay for higher occupied subbands is notably faster in some devices but not in others. We  analyze this using a tight-binding numerical model that includes the Zeeman, orbital and spin-orbit effects. When the first subband is spin polarized, we observe a dramatic suppression of supercurrent, which is also confirmed by the model and suggests an absence of significant triplet supercurrent generation.
\end{abstract}

\maketitle

\section{Introduction}
\textit{Context}. 
Semiconducor nanowire-based Josephson junctions have been explored as elements for superconducting transmon qubits~\cite{larsen2015semiconductor,de2015realization} and Andreev qubits~\cite{hays2021coherent}. 
The same one-dimensional(1D) super-semi hybrid system also fulfills basic requirements for emerging topological superconductivity and Majorana bound states (MBS) at nanowire ends~\cite{lutchyn2010majorana,oreg2010helical,frolov2020topological}. Attempts at exploring MBS in nanowire Josephson junctions were through the search for factional a.c Josephson effect~\cite{rokhinson2012fractional,bocquillon2017gapless,zhang2022missing}.
Topological qubits based on nanowire Josephson junctions containing MBS have been proposed theoretically~\cite{van2012coulomb,stenger2019braiding}. Subbands are created by quantum confinement of nanowires geometry, with the momentum components perpendicular to the nanowire axis as discrete and quantized values~\cite{lutchyn2011search}. In the simplest and purest form, Majorana modes are envisioned in a single-subband nanowire.

Mesoscopic electrical transport properties of supercurrents through quantum point contacts (QPC) ~\cite{beenakker1991josephson} were studied using atomic break junctions~\cite{muller1992conductance, goffman2000supercurrent}, and in two-dimensional electron gases~\cite{takayanagi1995observation, kjaergaard2016quantized, irie2014josephson}. InAs nanowires with transparent superconductor leads also feature few-subband transport~\cite{nishio2011supercurrent,goffman2017conduction,estrada2019charge,carrad2020shadow, abay2014charge}.

While studies of supercurrents in nanowires or point contacts have led to an array of interesting discoveries so far~\cite{rokhinson2012fractional,szombati2016josephson,spanton2017current,goffman2017conduction,zuo2017supercurrent,murani2017ballistic,ueda2019dominant,estrada2019charge,ligato2021preliminary,zhang2022evidence}, 
one outstanding challenge remaining is that supercurrent in the last occupied transverse mode (single subband) is strongly suppressed. It is either not observed or too weak to enable detailed studies
~\cite{nishio2011supercurrent,abay2014charge,irie2014josephson,estrada2019charge}. 
The reasons for this are not fully understood, however they are likely related to either finite interface transparency such as in devices involve ex-situ shell deposition, smaller induced gap such as in Al-InAs structures, residual scattering or other effects.

\begin{figure}
\includegraphics{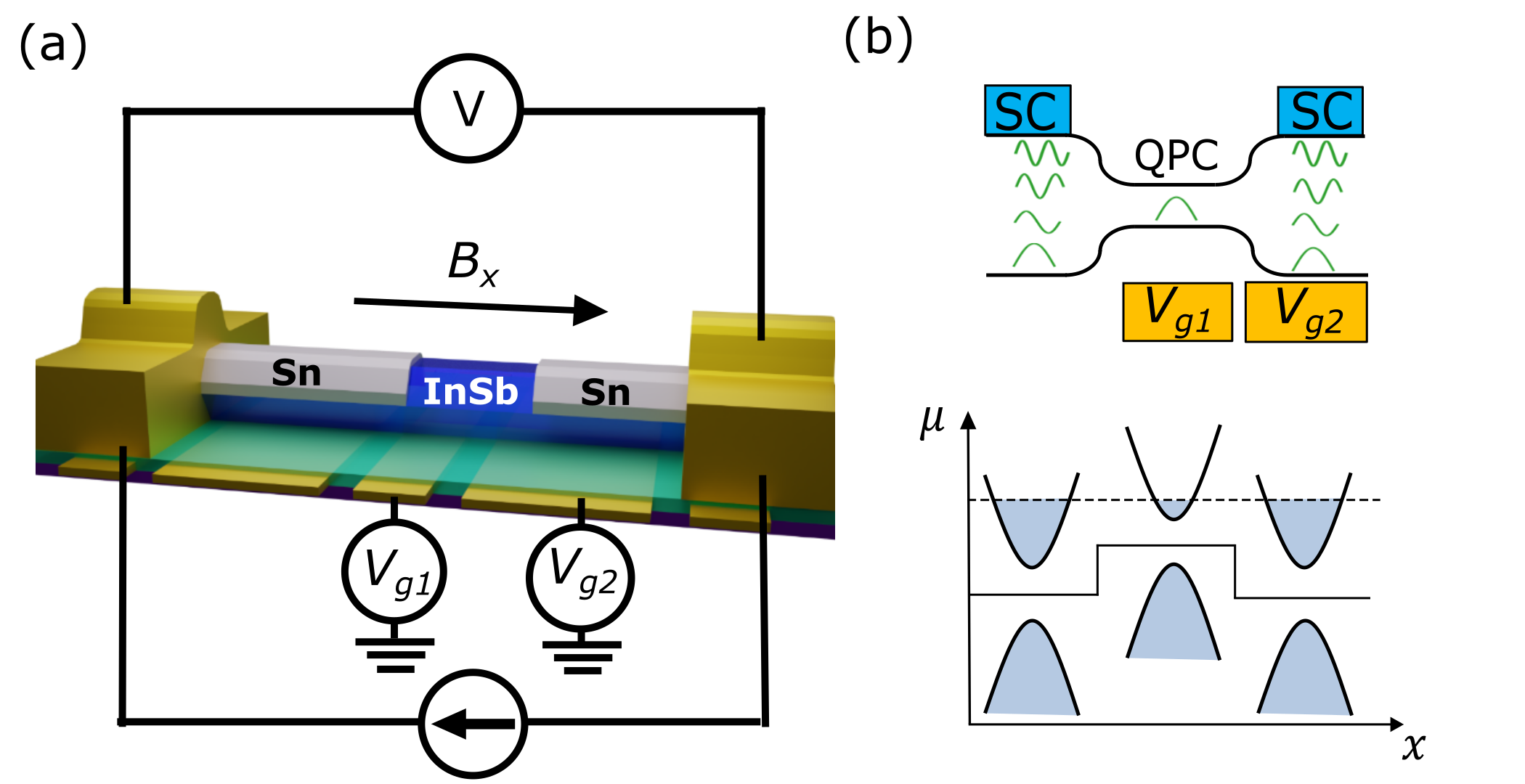}
\caption{\label{Fig1_device}
(a) Schematic of a shadow nanowire Josephson Junction device. The wire is along $\hat{x}$ which is also the direction of external magnetic field $B_x$. (b) Cartoon of a single-mode junction controlled by electrostatic gates $V_{g1}$ and $V_{g2}$. The leads are shown with higher subbands occupied. The lower panel corresponds to a chemical potential profile $\mu$(x) used in tight-binding simulations.}
\end{figure}

\textit{Approach.} Our approach is to combine InSb nanowires with Sn shells and study proximity-induced supercurrent in the nanwoires. Through advances in vapor-liquid-solid growth quasi-ballistic InSb nanowires were achieved~\cite{gul2015towards,badawy2019high}.  Quantized non-superconducting conductance has been established in InSb nanowires with normal metal electrodes~\cite{van2013quantized,kammhuber2016conductance}. The recently introduced superconducting Sn shells facilitate transparent contacts, and critical current-normal state resistance products exceeding temperature and those previously available in Al-based nanowire junctions~\cite{pendharkar2021parity}. Junctions in the Sn shell are defined on InSb nanowires by nanowire shadowing, which reduces processing and increases the likelihood of ballistic devices. Junction made with InSb nanowires and NbTiN leads are studied with same method, and the measurement results are compared with that in Sn-InSb devices as additional supplementary materials.

\textit{Results list}. 
 Subband-resolved transport is verified through the measurements of conductance at finite bias, and the evolution of it with gate voltage, source drain bias voltage and magnetic field. In the gate voltage and conductance range that corresponds to the first occupied transverse mode, we observe supercurrents as high as 20 nA. We investigate the gate voltage and magnetic field evolution of supercurrents in several nanowire devices both in the single-mode and in the multi-mode regimes. 
 The mechanisms of decay of supercurrent with magnetic field are studied including the relative contributions of orbital interference phenomena, residual disorder, spin-orbit and Zeeman effects by comparing the data to a tight-binding model. The spin-polarization of the first subband at finite field suppressed supercurrent dramatically, leading us to conclude that no triplet supercurrent was generated. 
 
\begin{figure}
\includegraphics{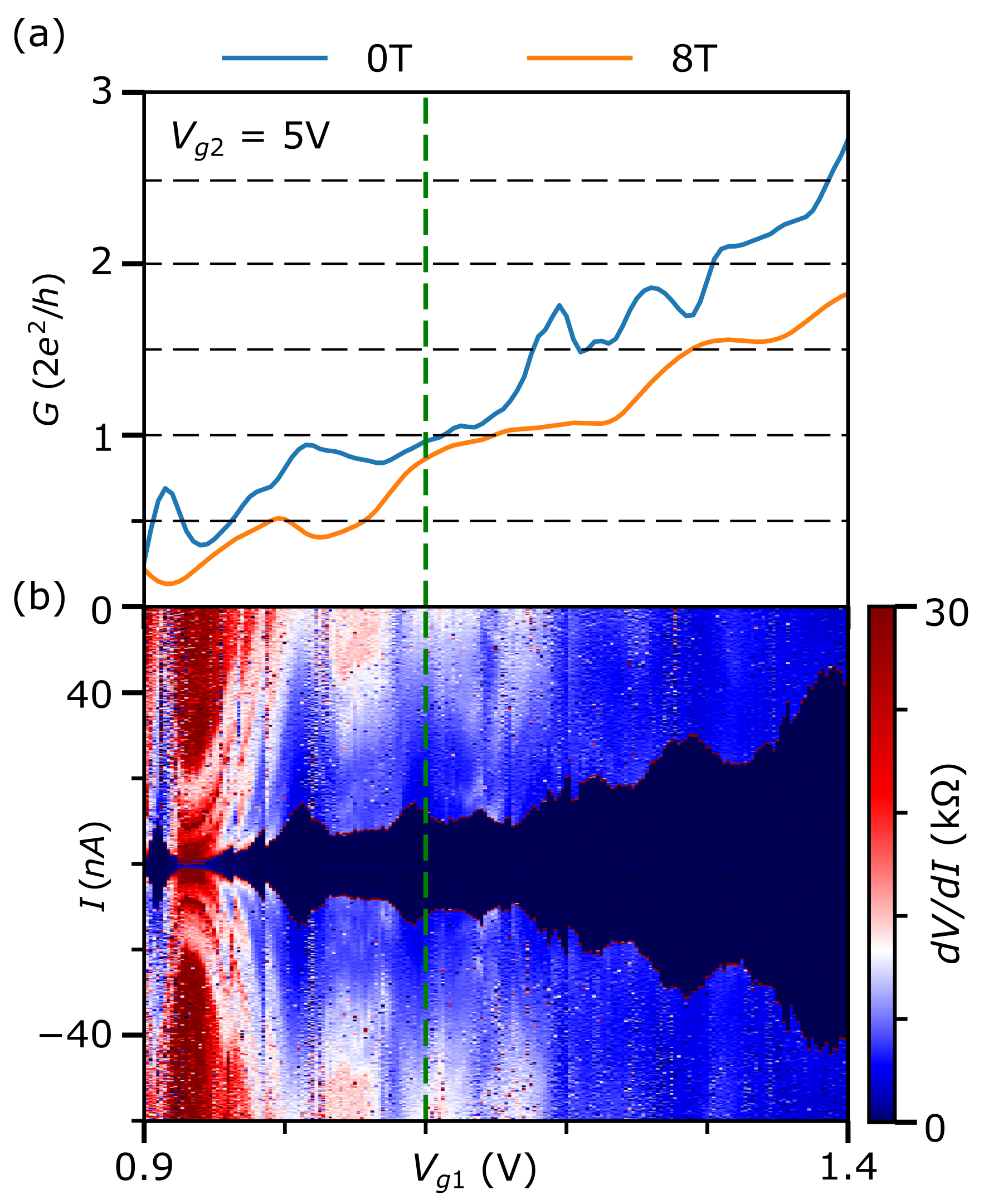}
\caption{\label{Fig2_1st_Ic}
(a) Differential conductance $G (dI/dV)$ taken at finite bias $V_{bias}$=2mV at B=0T and at zero bias for B=8T. (b) Current bias measurement of differential resistance $R (dV/dI)$ showing the evolution of supercurrent at B=0T. In this figure $V_{g2}$ = 5V. Green dashed line is used to indicate the approximate boundary between the first mode and higher modes based on conductance. Panels (a) and (b) are from separate measurements on Device A.
}
\end{figure}

\section{Figure 1: Device Description}
Fig.~\ref{Fig1_device}(a) presents a schematic diagram of the nanowire Josephson junction device. An InSb nanowire (blue) is half-covered by a Sn shell (silver) and positioned above local gate electrodes (gold), with Ti/Au contacts (gold). In order to prepare the junction itself, standing InSb nanowires approximately 100 nm in diameter are coated with a 15 nm layer of Sn~\cite{pendharkar2021parity}. In front of the nanowire, another nanowire shadows the flux of Sn to create two disconnected Sn segments. Nanowires with such shadow-defined junctions are transferred onto chips patterned with local electrostatic gate electrodes covered by HfOx dielectric. Contacts to wires are made using standard electron beam lithography and thin film deposition of Ti/Au layers. 

The supercurrent flows along the $\hat{x}$ direction, and an external magnetic field, $B_x$, is applied parallel to the wire. A current bias, $I_{bias}$, is applied across the device (illustrated by a black arrow), and the voltage across the device is measured using DC and AC multimeters in a two-point measurement setup. Two local bottom gates, $V_{g1}$ and $V_{g2}$, are located beneath the junction region. Measurements are performed in a dilution refrigerator with a base temperature of $\sim$50~mK equipped with a three-dimensional vector magnet.

Figure~\ref{Fig1_device}(b) uses cartoons to demonstrate the control of transverse mode numbers in the nanowire by local bottom gates during experiments, as well as the definition of local chemical potential in simulations. In the illustration, gate voltage $V_{g1}$ precisely adjusts the number of conduction channels, resulting in a single occupied transverse mode in the region labeled QPC (quantum point contact). Meanwhile, gate voltage $V_{g2}$ tunes one of the adjacent region which can have a higher subband occupancy. To emulate the realistic conditions in numerical simulations, we use two chemical potential $\mu$ settings, one for leads and one for the QPC.

\section{Figure 2: Supercurrent through the first electron mode}

In Fig.~\ref{Fig2_1st_Ic}(a) we plot two gate traces of conductance, one at zero magnetic field and one at large magnetic field (B=8T). The zero-field trace contains non-monotonic resonances due to quantum interference caused by backscattering, as well as charge jumps. This is in line with previous reports of quantum point contact behavior in nanowires~\cite{van2013quantized, kammhuber2016conductance}. Backscattering can be suppressed by large magnetic field, therefore the high magnetic field trace demonstrates a sequence of spin-resolved plateaus at $G_0/2 = 1\times e^2/h$ values. Using the high magnetic field trace as reference, we approximately identify the gate voltage interval that corresponds to the single occupied mode at B=0 ($V_{g1} <$ 1.1~$\mathrm{V}$, green dashed line), around conductance value of $G_0=2e^2/h$.  For the data shown, we estimate the highest conductance to be $6G_0$, corresponding to a maximum of 6 transverse modes. Comprehensive evidence of QPC behavior in device A is obtained from the magnetic field evolution of finite-voltage bias conductance maps for various gate voltage settings, which demonstrate diamond-shaped regions of relatively flat conductance in bias-gate space, and Zeeman splitting of these plateaus (see supplementary materials and Fig.\ref{Fig3_triplet_Ic}).

In Fig.~\ref{Fig2_1st_Ic}(b) we plot the current bias data that closely corresponds to the conductance traces from panel (a). The supercurrent appears as dark-blue regions around zero bias. On the left side of the green line, for more negative gate voltages, supercurrent is carried by the first transverse mode. The magnitude of the first mode switching current $I_{sw}$ in these and other data is from 10 to 20~$\mathrm{nA}$. Given the normal state resistance around 13~$\mathrm{k\Omega}$ (1$G_0$), the $I_{sw}R_N$ product falls within the range of 150-250~$\mu$eV(Fig.~\ref{figs_IcRn}). This value somewhat suppressed compared to the more open regime, but is of the same order as the superconducting gap gap of Sn ($\Delta$ = 650~$\mu$eV) and is consistent with values reported in previous studies~\cite{pendharkar2021parity, zhang2022evidence}. Thus signal levels allow for a deeper investigation of supercurrent in the few-subband regime.

\section{Figure 3: Suppression of supercurrent in the spin-polarized regime}

\begin{figure}
\includegraphics[width=0.98\columnwidth]{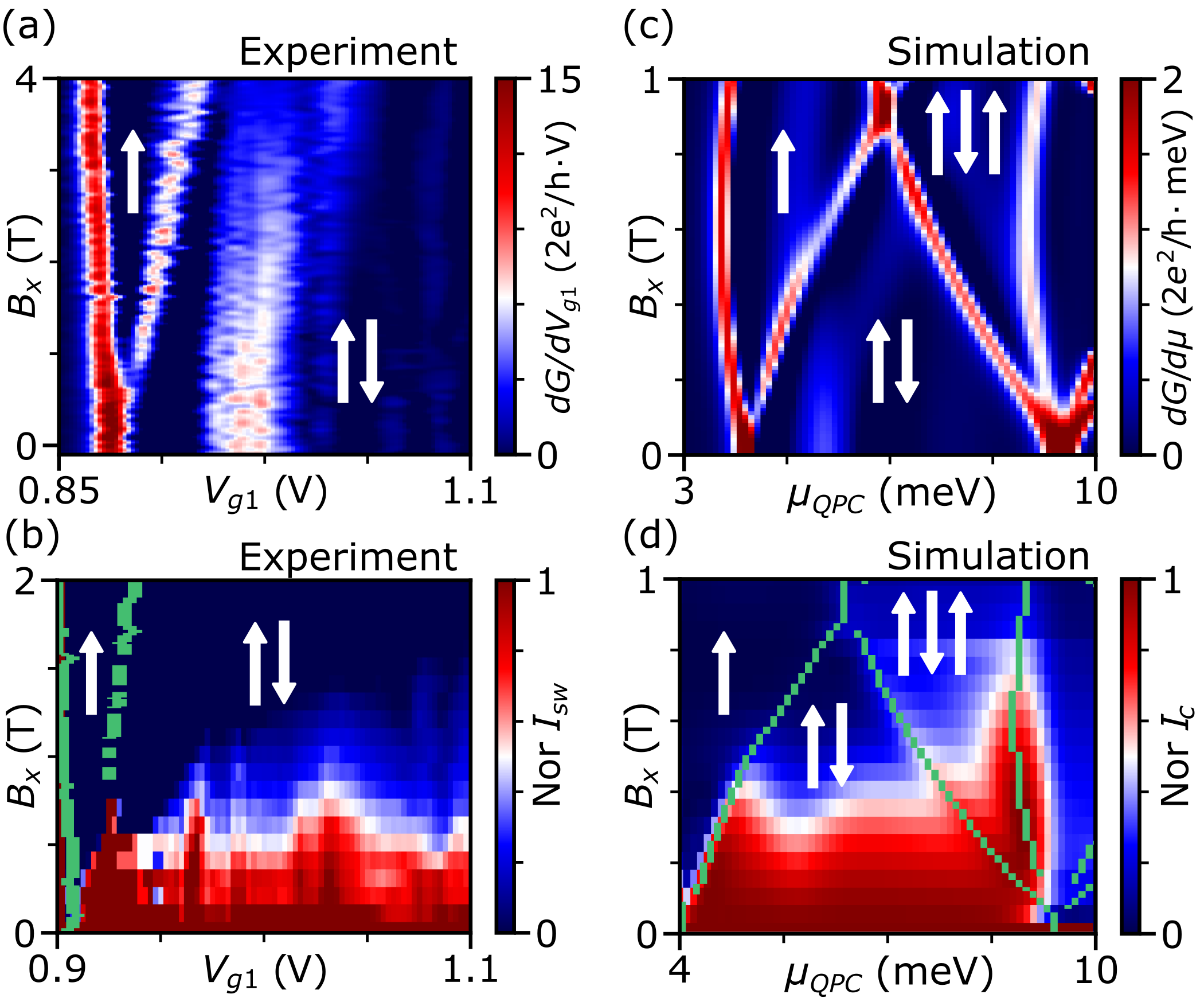}
\caption{\label{Fig3_triplet_Ic}
Comparison between experiment (a, b) and simulation (c, d): (a, c) Transconductance $dG/dV_{g1}$ vs $B_x$ and $V_{g1}$ ($\mu_2$) is measured at $V_{bias}$ = 2mV (0mV for simulation). $V_{g2}$ = 5~$\mathrm{V}$. The electron spin at each mode is indicated with white arrows. (b, d) Normalized $I_{sw}$/$I_c$ (Experiment/Simulation) is plotted as a function of $B_x$ and $V_{g1}$ ($\mu_2$). The boundary between the lowest spin-full mode and spin-polarized mode, extracted from (a), is indicated with a green line. The chemical potential under the leads, $\mu_{1},\mu_{3}$ = 20 $meV$, allows for a maximum of 6 available subbands through the leads. Other simulation parameters include spin-orbit strength $\alpha = 100 nm\cdot meV$, Zeeman g-factor ($g = 50$), effective mass $m_{eff} = 0.015m_e$, temperature $T = 100mK$, site disorder $\delta U = 0meV$. 
}
\end{figure}

The first question we investigate is the evolution of supercurrent as spin polarization develops in the QPC region, when the plateau $G_0/2 = 1\times e^2/h$ develops at finite magnetic field. The emergence of the plateau is demonstrated in Fig.~\ref{Fig3_triplet_Ic}(a) in the transconductance map - the characteristic "V"-shaped region corresponds to the spin-polarized plateau. Spin polarization becomes resolved at fields between 0.5-1.0T, while supercurrents are generally observed up to 2T. This allows for the study of the effect of spin polarization on supercurrent. 

Fig.~\ref{Fig3_triplet_Ic}(b) shows supercurrents extracted from gate voltage sweeps at fields between 0 and 2T. A switching current $I_{sw}$ is a current at which finite voltage develops across the junction, this may or may not closely follow the Josephson critical current which is a measure of Josephson coupling energy. In data processing, switching current is defined as $I_{bias}$ for which differential resistance exceeds 2$\mathrm{k\Omega}$. We extract the boundary of the spin-polarized region from panel (a) using peak finder scripts, and superimpose them with current evolution 2D map in panel(b) with green squares. Note that panel (a) is at $V_{bias}=2~mV$ while panel b is at zero voltage, meaning the boundaries may be shifted.  To quantify the decay rate of $I_{sw}$ in magnetic fields and compare it for different gate voltages, we normalize the magnitude of the switching current as $NorI_{sw} (B,V) = I_{sw}(B,V)/I_{sw}(B=0,V)$ and plot it as a function of $B_x$ and $V_g$.

Overall, the first mode supercurrent exhibits a decay up to $B_x$ = 1T, eventually disappearing as the system transits to the normal state. However, we found no supercurrent within the spin-up band, or at least the signal is rapidly suppressed in that region. At the same time, supercurrent found in the region of the phase diagram directly adjacent to the spin-polarized "V". This serves as an additional confirmation that it corresponds to the single subband regime with two spin channels.

\begin{figure*}
\includegraphics{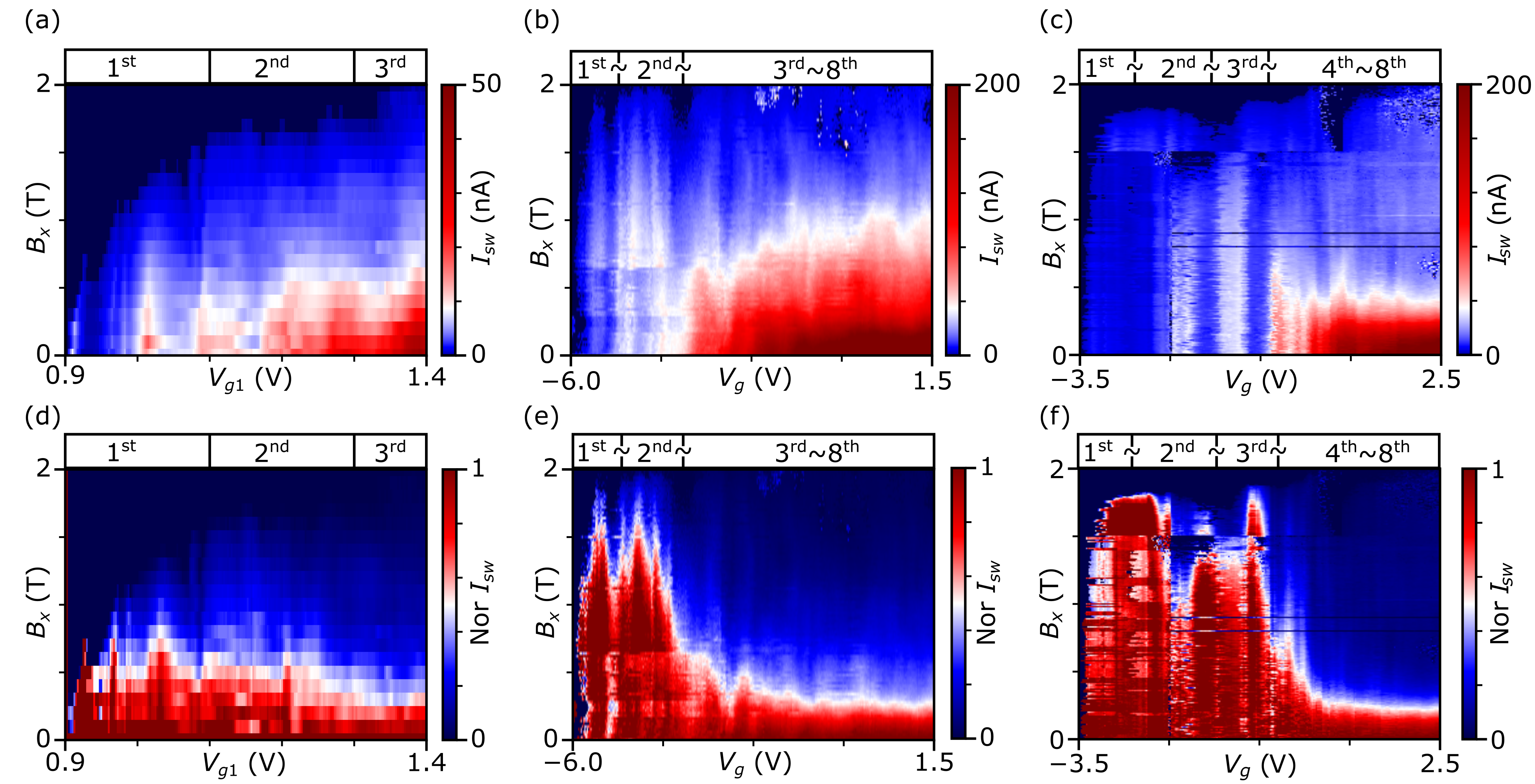}
\caption{\label{Fig4_Ic_map}
(a)-(c) Magnitude of switching current $I_{sw}$ extracted for gate scans and plotted as a function of gate voltages $V_g$ and parallel external field $B_x$ measured in Device A, B, and C. (d) Additional data for Fig.~\ref{Fig3_triplet_Ic}(b) (Device A) is presented, showcasing normalized $I_{sw}$ at higher gate voltages (more modes). (e) and (f): Normalized $I_{sw}$ as a function of $B_x$ and $V_g$ is measured in Device B and Device C. The boundary between different subbands are extracted at zero field and indicated by the yellow dashed line. The number of the highest transverse mode being occupied is labeled above each panel.}
\end{figure*}

We conduct a numerical investigation of the system's microscopic properties using a tight-binding model that mirrors the experimental geometry, as shown in Fig.\ref{Fig1_device}(b). This model was designed to study supercurrent interference in nanowires using KWANT~\cite{groth2014kwant, nijholt2016orbital, zuo2017supercurrent}. It includes the effects of spin-orbit interaction, the orbital vector potential, Zeeman splitting, electron temperature, and on-site disorder. To reproduce the quantized conductance results observed in our experiment, no disorder is considered here ($\delta U = 0meV$), same as what the initial version of the model did~\cite{zuo2017supercurrent}.

The new aspect here is the varied chemical potential along the nanowire, allowing for the possibility that the number of occupied subbands is not constant throughout the device. For the results in Fig.~\ref{Fig3_triplet_Ic}(c) we set chemical potentials in both the leads to $\mu_{lead} = 20meV$, corresponding to a few occupied subbands ($\approx$6), we then vary $\mu_{QPC}$. The boundary between the lowest spin-full and spin-polarized modes is extracted from the conductance calculation and plotted alongside the normalized $I_c$ in Fig.\ref{Fig3_triplet_Ic}(d) in a manner similar to how the experimental data are presented. The electron spin in each mode is labeled with white arrows. It should be noted that the values in the simulation do not exactly correspond to those in the experimental device but serve as indicators. For example, the boundary between $G_0$ and 1.5$G_0$ converges at $B_x = 0.9T$ in the simulation, whereas they never intersect in the experiment, even up to $B_x = 8T$. In the 1.5$G_0$ mode, we find that the supercurrent persists to a larger field compared to the first mode supercurrent, which is not observed experimentally but is likely due to microscopic details of the simulation.

In agreement with the experimental finding, the normalized $I_c$ does not survive in the spin-polarized regime at 0.5$G_0$. This indicates that the model correctly captures the system and does not predict triplet supercurrent under the most basic conditions. Experiments on superconductor-ferromagnet-superconductor structures found that in junctions with pristine interfaces supercurrent vanishes in shorter junctions, while counterintuitvely it survives to in longer junctions when interfaces are disordered~\cite{frolov2008imaging}. A disordered interface is believed to facilitate spin flipping into triplet state and flipping back to singlet at the second interface. It is worth investigating if interface roughness, in combination with spin-orbit interaction and/or magnetic impurities culd extend supercurrents into the spin-polarized regime in superconductor-semiconductor junctions, hinting at the generation of triplet supercurrents.

\section{Figure 4: single-mode versus multi-mode regimes}

Earlier work on nanowires suggested that the single-mode supercurrents are unique, because without other occupied subbands there is no inter-subband interference. This interference was associated with rapid decay of supercurrents in magnetic field, and therefore a slower decay is expected for a single-subband junction~\cite{zuo2017supercurrent}.

In Fig.~\ref{Fig4_Ic_map} we show switching current maps in gate-field space from three devices. Device A is the source of data in Figs.~\ref{Fig2_1st_Ic} and ~\ref{Fig3_triplet_Ic}, where the few mode regime was carefully explored. Devices B and C are fabricated using the same Sn-InSb nanowires and in the same geometry (see supplementary materials for device images and additional data). While conductance steps are observed, the QPC evidence is less comprehensive in devices B and C. On the other hand, B and C are studied into the more open regime. The number of occupied subbands is estimated from conductance and indicated above the figures.

In all three junctions, supercurrents grow at more positive gate voltages that correspond to higher subband occupations [Figs.~\ref{Fig4_Ic_map}(a)(b)(c)]. In magnetic field, signal is observed up to, and occasionally beyond B=2T. The magnetic field decay rate of switching current can be explored in maps normalized to zero-field values [Figs.~\ref{Fig4_Ic_map}(d)(e)(f)]. Devices B and C clearly exhibit a more rapid relative decay of switching current in the multi-mode regime, compared with the 1-2 mode regime. The rapid decay takes place at fields below approximately 0.5T followed by a persistent lower signal. In the single-mode regime, no rapid decay is observed, yet the overall signal level is smaller and comparable to that seen at higher fields in the multi-mode regime (see supplementary information for detailed magnetic field dependences of current-voltage characteristics for all three devices.) Results in device A do not extend to more modes, but from the data available they are not in contradiction with findings from B and C.

\section{Alternative explanations for magnetic field decay rates}

\begin{figure}
\includegraphics[width=0.9\columnwidth]{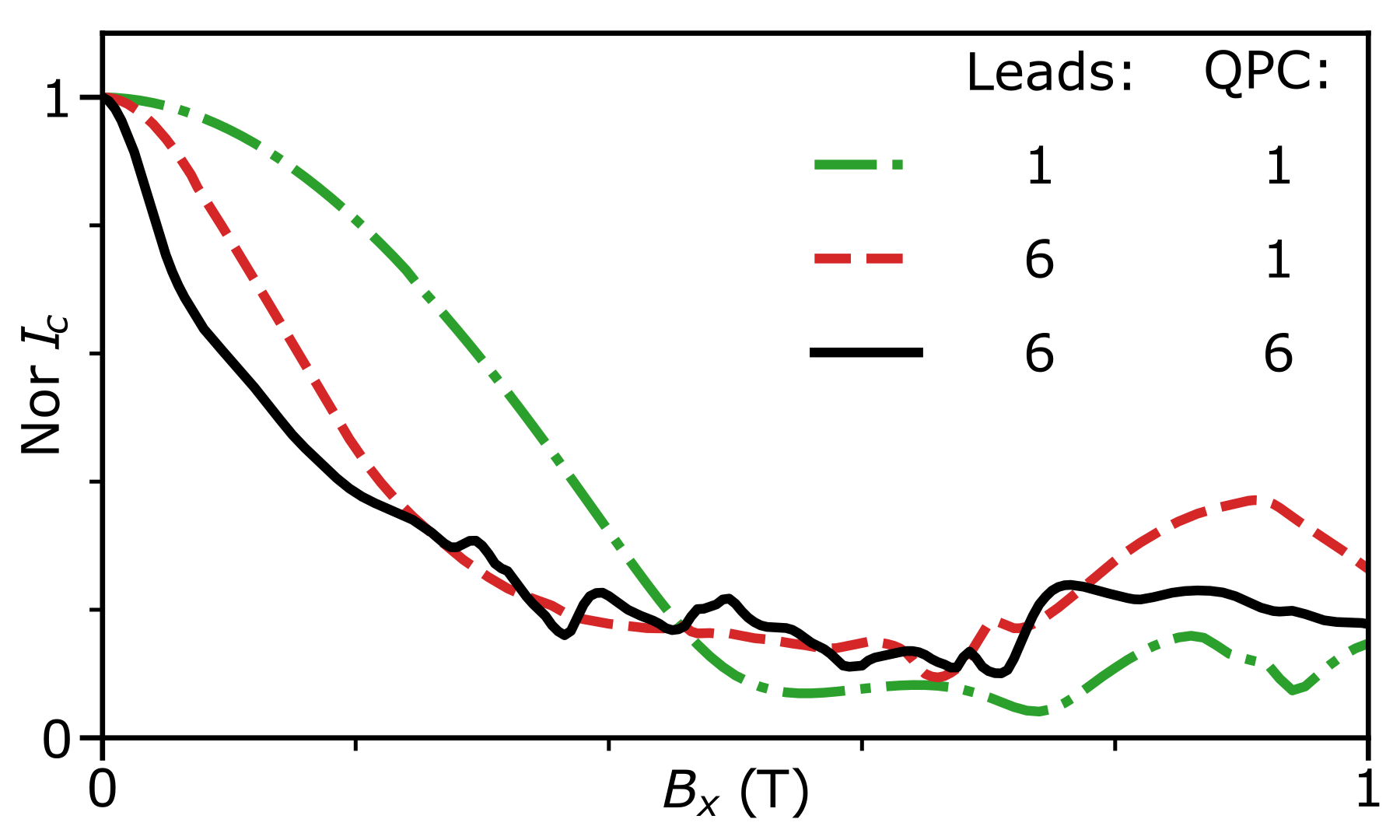}
\caption{\label{Fig5_Ic_sim}
Normalized critical current as a function of parallel field $B_x$ for different combinations of terms in the Hamiltonian of the numerical model. The Zeeman effect ($g = 50$) and spin-orbit strength $\alpha = 100 nm\cdot meV$ are present in all curves. The local chemical potential and corresponding number of conduction channels in the leads and QPC region are illustrated with a small diagram in each panel. $\mu=$ 5 and 20~$\mathrm{meV}$ result in one and six spin-full transverse modes in the nanowires, respectively. The other simulation parameters are consistent with those used in Fig.~\ref{Fig3_triplet_Ic} (b) and (d).
}
\end{figure}

We discuss the decay rates of supercurrent in magnetic field from the experimental and theory points of view. On the experiment side, one concern that should be mentioned is that smaller switching currents can deviate substantially from the actual Josephson critical currents. This concern is shared by all studies of switching currents on mesoscopic junctions. We note that, in contrast with many studies, here the switching into the normal state remains sharp in the regime of interest, at least near zero applied magnetic field.  Another experimental concern is that the shapes of $I_{sw}(B)$ can be affected by presence of finite voltage resonances such as due to Multiple Andreev Reflections (MAR) frequently observed in these junctions leading to peculiar non-monotonic field dependences (see supplementary information for examples, and future work for an in-depth study.) 

\section{Figure 5: Tight-binding simulation of junction with leads}

On the theoretical side, we started with a basic assumption of only a single mode occupied in the device. While this may be the case in the middle of the junction, the leads may have higher subband occupations. Interference between supercurrents carried by different subbands can take place in the leads and result in faster decay. The earlier model did not include this consideration, so we address it here. 

Using KWANT, we numerically calculate the evolution of $I_c$ in the parallel field $B_x$ [Fig.\ref{Fig5_Ic_sim}]. The chemical potential in the leads and junction region is locally adjusted to 5~$\mathrm{meV}$ (one mode) or 20~$\mathrm{meV}$ (six modes). Spin-orbit interaction and Zeeman effect are present in all calculations. The effects of small disorder are explored in the supplementary information. In agreement with earlier results~\cite{zuo2017supercurrent}, the model demonstrates the slowest decay when both the leads and the QPC region are set to one occupied subband. The decay is more rapid when the leads are set to 6 subbands, and opening the QPC to 6 subbands accelerates the decay further. The difference between QPC and 1 and at 6 is not dramatic when the leads are open.

Experimentally it is much more challenging to realize a device where the entire nanowire is in the single subband regime, compared with realizing a short QPC region. Supplementary information shows more maps for different settings of $V_{g2}$ for device A. More negative settings of $V_{g2}$ result, in principle, in a lower subband occupation in one of the leads. While the decay rates become more uniform at negative $V_{g2}$, the limited range of $V_{g1}$ and the inability to tune the second lead region prevent us from concluding on the origins of this effect and whether the numerical simulations explain it. Since single-mode nanowires are desirable for Majorana zero mode experiments, future work should focus on realizing this regime and the insights obtained in the present work may be helpful.

\section{Acknowledgements}
We thank G. Badawy, S. Gazibegovic, E. Bakkers for providing InSb nanowires. We thank R. Mong for discussions. We acknowledge the use of shared facilities of the NSF Materials Research Science and Engineering Center (MRSEC) at the University of California Santa Barbara (DMR 1720256) and the Nanotech UCSB Nanofabrication Facility.

\section{Funding}
Work supported by the NSF PIRE:HYBRID OISE-1743717, NSF Quantum Foundry funded via the Q-AMASE-i program under award DMR-1906325, U.S. ONR and ARO and France ANR through Grant No. ANR-17-PIRE-0001 (HYBRID).

\section{Data Availability}
Curated library of data extending beyond what is presented in the paper, as well as simulation and data processing code are available at~\cite{zenodo_qpc}.

\section{Duration and Volume of Study}
This project was started in June 2018 and ended in December 2022. The simulation analysis phase was completed in March 2023. Within this report, a total of six devices were studied. Two of these devices were fabricated using pure InSb nanowires, with ex-situ deposited NbTiN leads, and were fabricated in 2018. The remaining four devices were fabricated using nanowires that are reported in Ref.~\cite{pendharkar2021parity}. For the purpose of this project, 59 devices across 8 chips were fabricated and measured during 11 cooldowns in dilution refrigerators. These measurements yielded approximately 5900 datasets, of which around 2700 datasets were shared with the project reported in Ref.\cite{zhang2022evidence}.

\section{References}

\bibliographystyle{apsrev4-1}
\bibliography{references.bib}

\clearpage
\beginsupplement

\begin{center}
    \textbf{\large{Supplementary Materials}}
\end{center}

\section{Fabrication and Measurements} \label{Sec_fab_measure}
\textbf{Sn-InSb devices}
Nanowires are transferred onto a silicon (Si) chip with predefined local gates. These electrostatic local gates are patterned using 100 keV Electron Beam Lithography (EBL) on undoped Si substrates. The local gates feature mixed widths of 80 and 200~$\mathrm{nm}$ and are separated by a distance of 40~$\mathrm{nm}$. The gates are metalized through electron beam evaporation of 1.5/6~$\mathrm{nm}$ Ti/PdAu and subsequently covered by a 10~$\mathrm{nm}$ ALD HfOx dielectric layer.

According to the Scanning Electron Microscope images of all devices examined in this report, the lengths of the Josephson junctions (JJs) created with Sn-InSb nanowires range from 120 to 150~$\mathrm{nm}$. The junction width is consistent with the nanowire width, which is approximately 120~$\mathrm{nm}$. After positioning the nanowires onto the gates, the entire chip is coated with PMMA 950 A4 electron beam resist. The resist is dried at room temperature using a mechanical pump in a metal desiccator for 24 hours. EBL is then employed to define normal lead patterns. Following development, resist residue is cleaned in an oxygen plasma asher. Within the electron beam evaporator, an in-situ ion mill is initially used to remove the AlOx capping layer from the nanowires in the contact area, after which 10~$\mathrm{nm}$/130~$\mathrm{nm}$ Ti/Au is deposited on the chips.

Two-point transport measurements are conducted using a current source and a parallel voltage measurement model, with multiple filtering stages placed at different temperatures.  First, the measurement circuit uses a dedicated electrical ground not connected to interference sources such as computers, fire alarms, pumps, lights etc. At room temperature, we use pi-filters, then the signal is passed through charcoal clad twisted pair shielded wires that are 1.5 meters long, this provides attenuation of signals at 1-10 GHz frequencies and beyond.  Then, at the mixing chamber plate, the wires enter a copper shield through a copper-powder filter. The shield around the sample is surrounded by several radiation shields that are part of the dilution refrigerator standard setup. Finally, inside the copper can we place another two-stage RC filter on each line. This, as well as bespoke low-noise amplifiers, allow us to measure signals in the range of 10 nV and 100 fA in ideal conditions. Series resistance from the filter and the measurement model in $V_{source}$-$I_{measure}$ scan is $R_{in}$ = 7.04~$\mathrm{k\Omega}$. In $I_{source}$-$V_{measure}$ scan, $R_{in}$ = 4.04~$\mathrm{k\Omega}$.

\begin{figure}[H]
\includegraphics{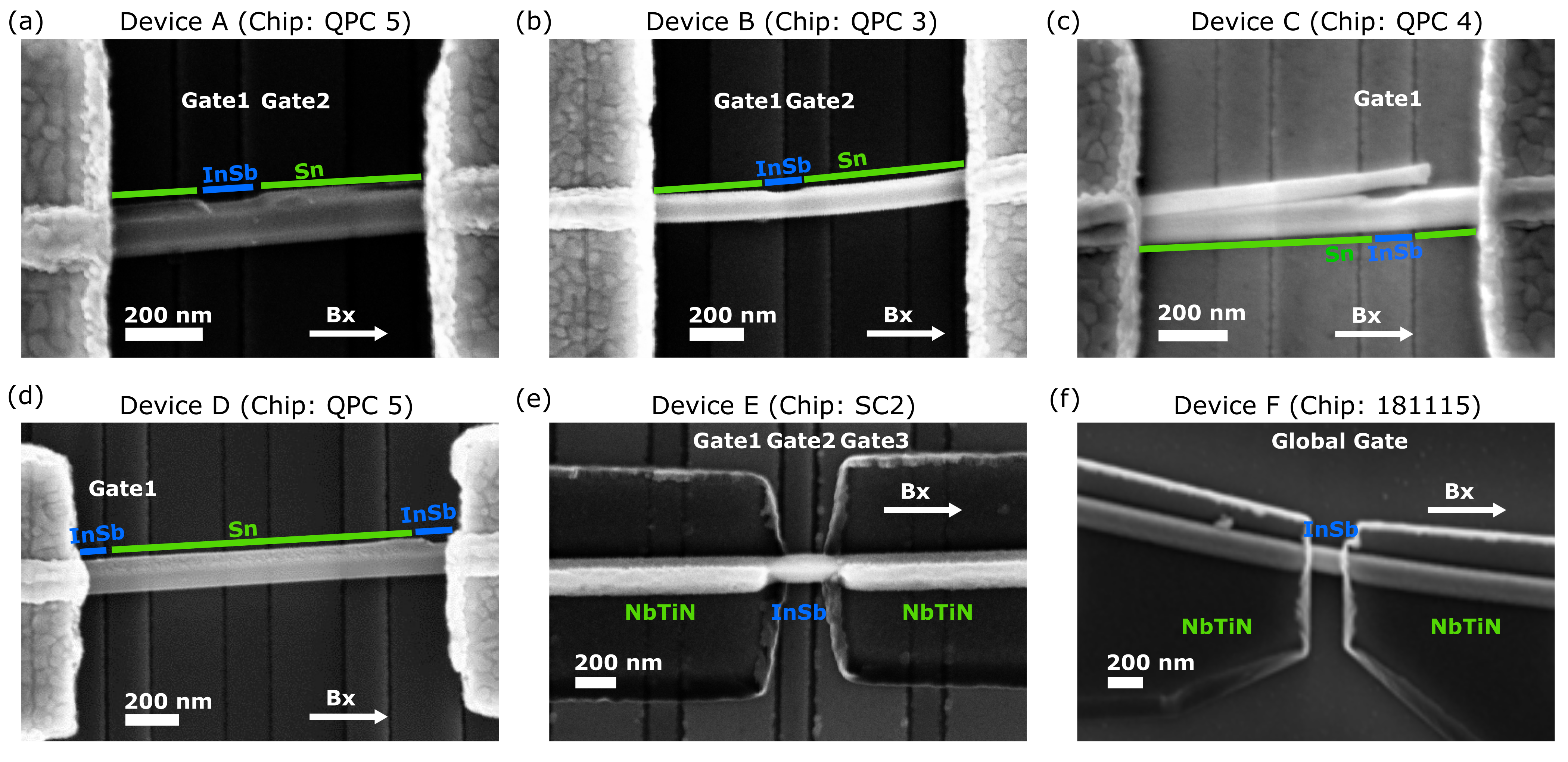}
\caption{\label{figs_devices} Scanning electron microscope (SEM) images of devices measured in reports, listed as Device name and (Chip name). Sn-InSb devices: (a) Device A (QPC5). (b) Device B (QPC3). (c) Device C (QPC4). (d) Device D (QPC5). NbTiN-InSb devices: (e) Device E (SC2). (f) Device F (181115). The direction of applied magnetic field $B_x$ is indicated with white arrows.}
\end{figure}

\textbf{NbTiN-InSb devices}
In Device E (SC2), InSb nanowires are transferred onto a Si chip with predefined local bottom gates(Fig.~\ref{figs_devices} (e)). The chip shares the same geometry as those used for Sn-InSb devices. Device F (181115) is fabricated using highly-doped Si chips that employ the entire substrate as global bottom gates(Fig.~\ref{figs_devices} (f)). The global gate chip is covered by a 285nm SiOx dielectric layer.

After transferring the nanowires, the entire chip is coated with PMMA 950 A4 electron beam resist and baked for 15 minutes at 175~$^{\circ}$C to dry out the resist. The superconducting contacts are defined by EBL. Following development, a sulfur passivation technique is used to remove the oxide layer from InSb nanowires~\cite{suyatin2007sulfur}. In the film deposition sputtering system, a smooth in-situ ion mill with argon plasma is utilized to eliminate sulfur and resist residue for 10 seconds at 15~$\mathrm{W}$, after which 10~$\mathrm{nm}$/140~$\mathrm{nm}$ NbTi/NbTiN is deposited onto the chips. Unwanted patterns are removed by soaking in acetone overnight and performing lift-off with a pipette to blow away the resist.

According to the SEM images, the length of the JJs in Device E and F are approximately 200~$\mathrm{nm}$. The junction width is consistent with the nanowire width, and is similar to the Sn-InSb devices. 

\hfill

\section{Full data of Device A}
\subsection{Voltage-current characteristic}
\begin{figure}[H]
\includegraphics{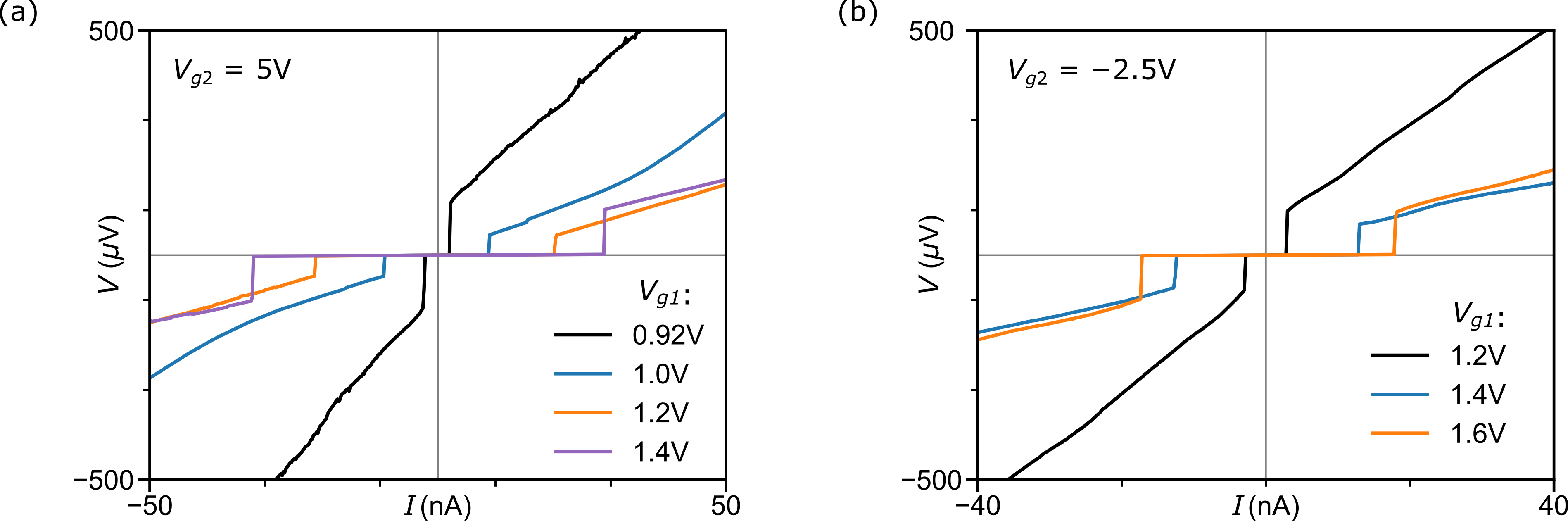}
\caption{\label{figs_QPC5_IV} Several voltage-current characteristic for device A taken at different QPC gate voltage $V_{g1}$. The curve are derived from Fig.~2(b) from the main text for lead gate $V_{g2}$ = 5~V (a) and Fig.~\ref{figs_QPC5_Ic_1st}(b) for $V_{g2}$ = -2.5~V (b).}
\end{figure}

Characterization of voltage-current relation of device A at $B$ = 0~T is shown in Fig.~\ref{figs_QPC5_IV}. Taken known series resistance $R_{in}$ = 4.44~k$\Omega$ into account. The series resistance is known being caused by the measurement set up (filters in the series and measurement module), and the total value is confirmed by a shorted connection in the dilution refrigerator that being measured at the same time.
\clearpage

\subsection{Evidence of Quantum Point Contacts}
\begin{figure}[H]
\includegraphics{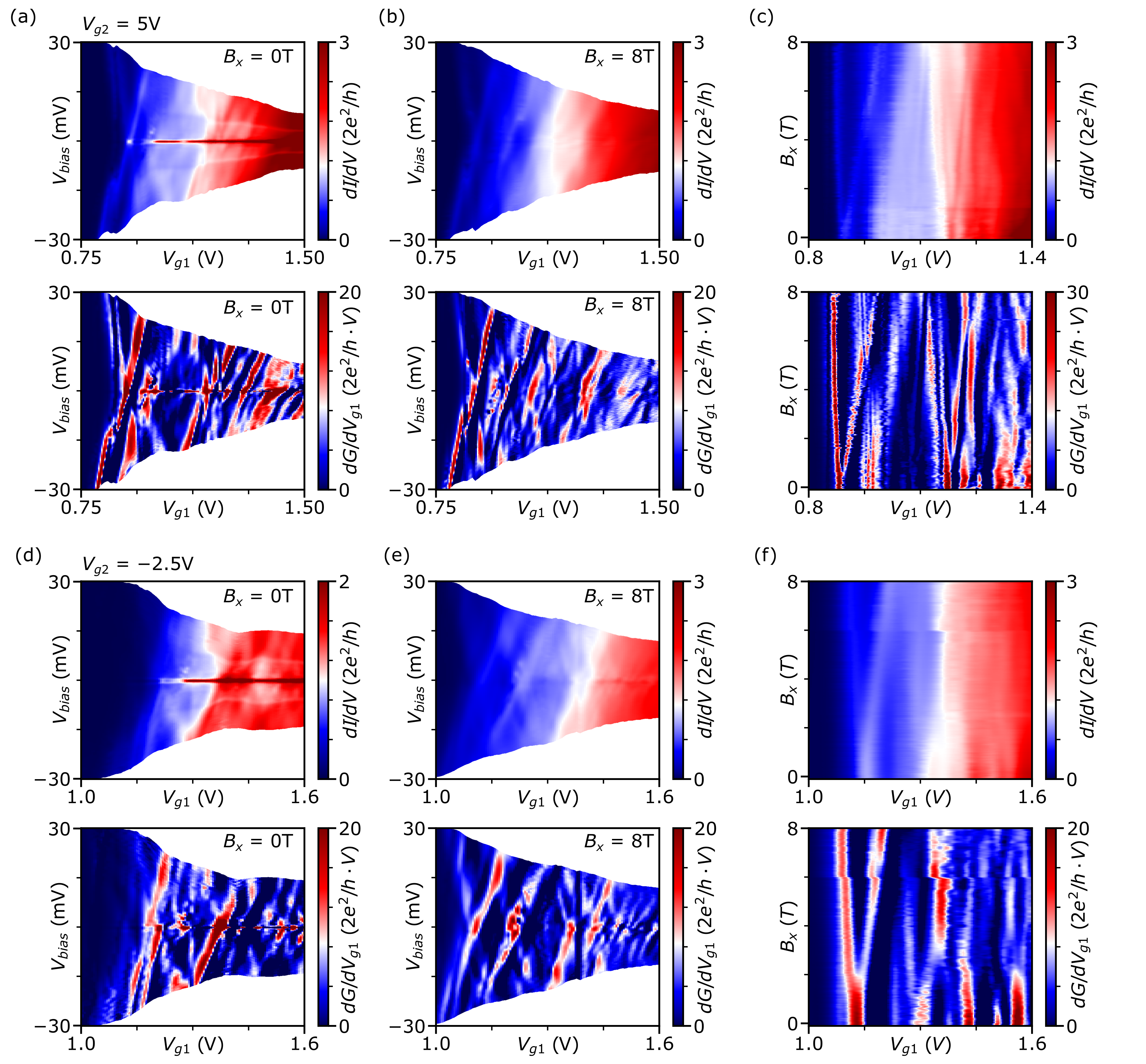}
\caption{\label{figs_QPC5_QPC} (a) Upper: Differential conductance $G$ = $dI/dV_{bias}$ as a function of junction gate voltage $V_{g1}$ and voltage across the device $V_{bias}$ measured at lead gate voltage $V_{g2}$ = 5~$\mathrm{V}$ and external field $B_x$ = 0~$\mathrm{T}$, a series resistance $R_{in}$ = 8.7k~$\Omega$ is used to calculate the conductance and extract from $Vbias$ while plotting. Lower: Transconductance $dG/dV_{g1}$ as function of $V_{g1}$ and $V_{bias}$ plotted with same data to show the boundary of conductance plateau. (b) Similar as (a) but external field $B_x$ = 8~$\mathrm{T}$. (c) Upper: $G$ as a function of $V_{g1}$ and $B_x$. Lower: Full data of Fig.~\ref{Fig3_triplet_Ic}(a), transconductance $dG/dV_{g1}$ as function of $V_{g1}$ and $B_x$. (d)-(f) similar as (a)-(c) but the lead gate voltage $V_{g2}$ = -2.5~$\mathrm{V}$.}
\end{figure}

The devices initially undergo characterization by sweeping the gate voltages ($V_g$) at a fixed voltage bias across the device ($V_{bias}$), while concurrently measuring the current ($I$) passing through the device as a function of $V_g$. The series resistance $R_s$, used for calculating the differential conductance, comprises two components: $R_in$ from various filtering stages (refer to Sec.~\ref{Sec_fab_measure}), and $R_contact$ at the Au-Sn and Sn-InSb interface. The magnitude of $R_s$ is derived from the $I$-$V_g$ trace, where $R_s$ equals $V_{bias}$ (10$\mathrm{mV}$) divided by $I_{max}$. When the current no longer increases with rising gate voltage, it implies that all conduction channels in the nanowire junction are fully open, signifying device saturation. At this point, only the series resistance acts as a transport barrier.

Evidence of quantum point contacts (QPC) construction within our nanowire Josephson junction is demonstrated by plotting the differential conductance $G$ ($dI/dV$) against the voltage across the device ($V_{bias}$) and the junction gate voltage ($V_{g1}$), with an external field $B_x$ = 0~$\mathrm{T}$/8~$\mathrm{T}$ and lead gate voltage $V_{g2}$ = 5~$\mathrm{V}$ (see Fig.~\ref{figs_QPC5_QPC} (a) and (b)). We used a series resistance $R_s$ = 8.7~$\mathrm{k\Omega}$ to calculate the differential conductance, subtracting its voltage drop contribution from $V_{bias}$ during plotting. Comparison reveals a diamond-shaped region with quantized conductance in units of 2$\mathrm{e^2/h}$ (1$G_0$) at 0~$\mathrm{T}$, alongside an extra region with conductance corresponding to 1$\mathrm{e^2/h}$ (0.5$G_0$) at 8~$\mathrm{T}$ and a smaller $G_0$ region. Transconductance $dG/dV_{g1}$ over the junction gate voltage marks each quantized conductance region's boundary with red lines.

Fig.~\ref{figs_QPC5_QPC}(c) illustrates a scan of $G$ as a function of $B_x$ and $V_{g1}$ at a fixed voltage bias $V_{bias}$ = 2~$\mathrm{mV}$. Here, we observe the emergence of a spin-polarized regime ($G$ = 0.5$G_0$) due to Zeeman splitting. To account for the Josephson current at zero voltage bias and Andreev states at a voltage bias smaller than 2$\Delta$ potentially resulting in higher conductance than the normal state conductance corresponding to the conduction mode, $V_{bias}$ is set higher than 2$\Delta$, where $\Delta$ is the gap of Tin (approximated as 650~$\mathrm{\mu eV}$). A segment of the transconductance region from the Zeeman effect scan is used in Fig.~\ref{Fig3_triplet_Ic} in the main text to differentiate between the spin-degenerate and spin-polarized regions.

In Fig.~\ref{figs_QPC5_QPC} (d) to (f), we replicate the same scans as in panel (a) to (c), but with a different lead gate voltage setting $V_{g2}$ = -2.5~$\mathrm{V}$. We discovered that the lead gate voltage value does not influence the presence of the quantized conductance region and the Zeeman effect. In Fig.~\ref{figs_QPC5_QPC_gvsg}, we present additional conductance plots at varying lead gate voltages, thus demonstrating that the QPC in the junction region remains unaffected by changes in the lead gate voltage $V_{g2}$. This finding highlights the stability of our constructed QPCs within the nanowire Josephson junctions, regardless of variations in lead voltage settings.
\clearpage

\subsection{Gate dependence of Conductance}
\begin{figure}[H]
\includegraphics[width = \textwidth]{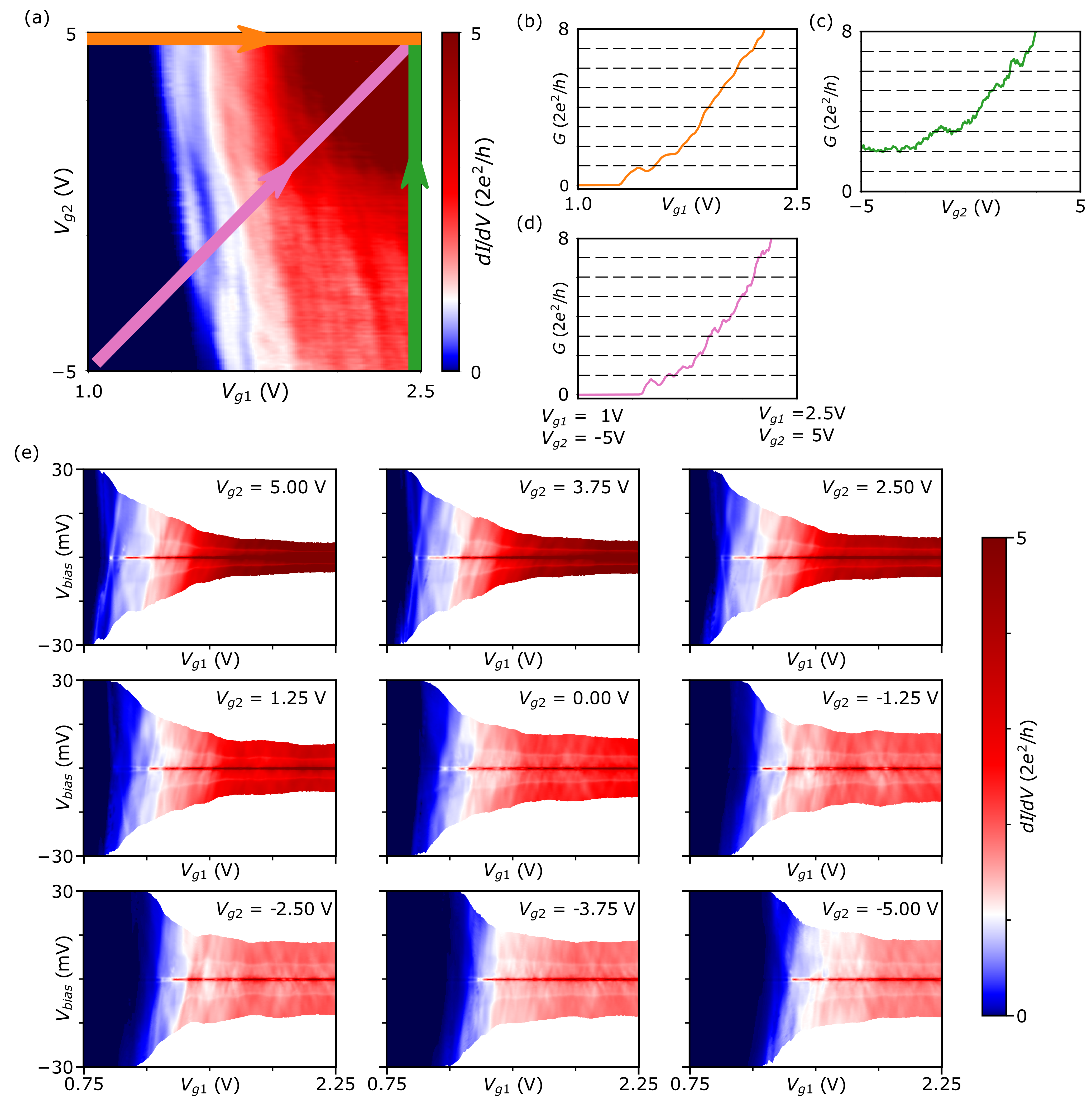}
\caption{\label{figs_QPC5_QPC_gvsg} (a) Differential conductance $dI/dV$ as a function of junction gate $V_{g1}$ and contact gate $V_{g2}$, measured with fixed voltage bias $V_{bias}$ = 2mV and external field $B_x$ = 0T. (b)-(d) Three line cuts taken from (a): (b) Conductance as a function of $V_{g1}$ at $V_{g2}$ = 5V. (c) Conductance as a function of $V_{g2}$ at $V_{g1}$ = 2.5V. (d) Conductance as a function of $V_{g1}$ and $V_{g2}$ when both gates are varying. (e) $G$ as a function of $V_{g1}$ and $V_{bias}$ at $B_x$ = 0T and a series of different $V_{g2}$.}
\end{figure}

In Fig.~\ref{figs_QPC5_QPC_gvsg}, we explore the gate dependence in Device A. According to Fig.~\ref{figs_QPC5_QPC_gvsg} (a), the lead gate voltage $V_{g2}$ exhibits a relatively minor effect in modulating the conduction channels and cannot independently close the nanowires within the measurable range of -5~$\mathrm{V}$ to 5~$\mathrm{V}$. However, the gate voltage range of $V_{g1}$, where only a single mode is available in the junction, is still influenced by the lead gate $V_{g2}$. This aspect is taken into account when analyzing the impact of the orbital and Zeeman effects on the supercurrent across different modes.

In Fig.~\ref{figs_QPC5_QPC_gvsg} (e), we present a conductance scan as a function of $V_{bias}$ and $V_{g1}$ at varying lead gate voltages $V_{g2}$, but with a fixed external field $B_x$ = 0~$\mathrm{T}$. We observe distinct diamond-shaped conductance plateaus across all $V_{g2}$ values. This serves as evidence that the lead gate does not disrupt the ballistic transport and electron confinement within the nanowire junction region.

\subsection{Extended Normalized $I_{sw}$ map}
\begin{figure}[H]
\includegraphics{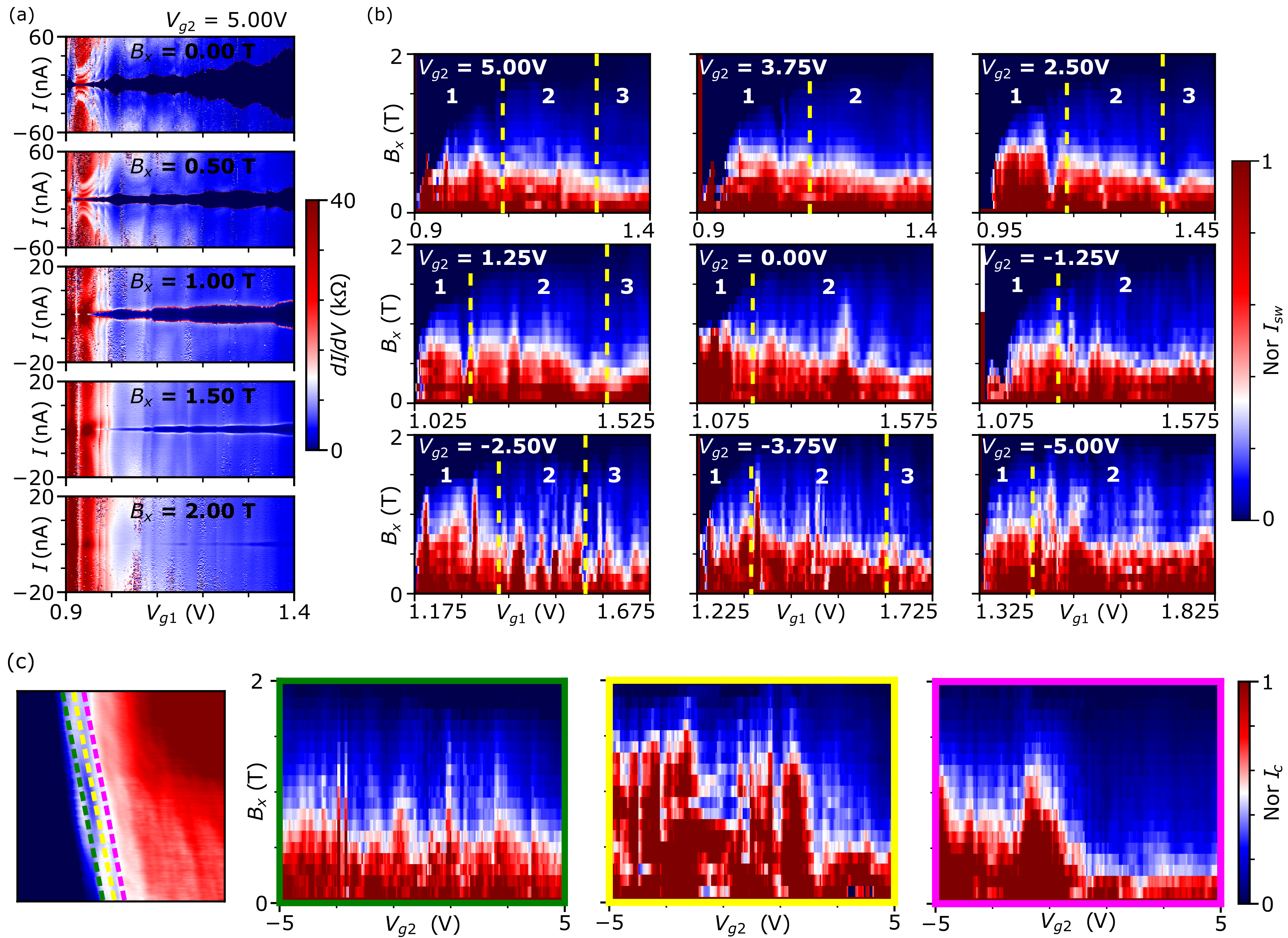}
\caption{\label{figs_QPC5_Ic_map} (a) Gate scan of supercurrent at a series of external $B_x$, the lead gate voltage $V_{g2}$ is 5~$\mathrm{V}$. (b) Normalized $I_{sw}$ as a function of junction gate $V_{g1}$ and external field $B_x$, the voltage applied to $V_{g2}$ is labeled in each panel. The boundary between how many modes are occupied in the junction region is indicated with yellow dashed line and the number of conduction channels is labeled with white numbers. The number of occupied modes is derived from Fig.~\ref{figs_QPC5_QPC}. (c) Normalized $I_{sw}$ map as a function of $B_x$ while both gate voltages $V_{g1}$ and $V_{g2}$ are varying to along three dashed line in the gate versus gate scan adapted from Fig.~\ref{figs_QPC5_QPC_gvsg} (a) to stay in the first mode. The color of the square box is corresponding to the voltage tuning trace dashed line.}
\end{figure}

The effects of junction gate voltage on superconductivity in the presence of an external field are discussed in Fig.~\ref{Fig4_Ic_map} and explored via simulations that consider both the chemical potential in the junction and the lead regions. These simulations are depicted in Fig.~\ref{Fig5_Ic_sim} in the main text. We conclude that the orbital effect is the primary factor leading to the suppression of the Josephson current, and the chemical potential under the leads also influences superconductivity, even when the conduction channels in the nanowires remain the same. Here, we present extended data to further substantiate our interpretation.

In Fig.~\ref{figs_QPC5_Ic_map} (a), we provide several raw data sets from gate scans of the Josephson current at various external field strengths. The deep blue represents the zero resistance state when a current bias is applied across the device. We observe that the switching current $I_{sw}$ at higher gate voltages decays more rapidly compared to the $I_{sw}$ at lower gate voltages. The $I_{sw}$ at different gate voltages does not maintain a consistent magnitude at high fields, illustrating that the supercurrent is continuously tunable by the gate voltage. This suggests that it is not contributed by a continuous shell but by the nanowire junction. These types of gate scans are the source of all $I_{sw}$ maps presented in this paper. We extract the magnitude of $I_{sw}$ and plot them as functions of $B_x$ and $V_g$.

\begin{figure}[H]
\includegraphics{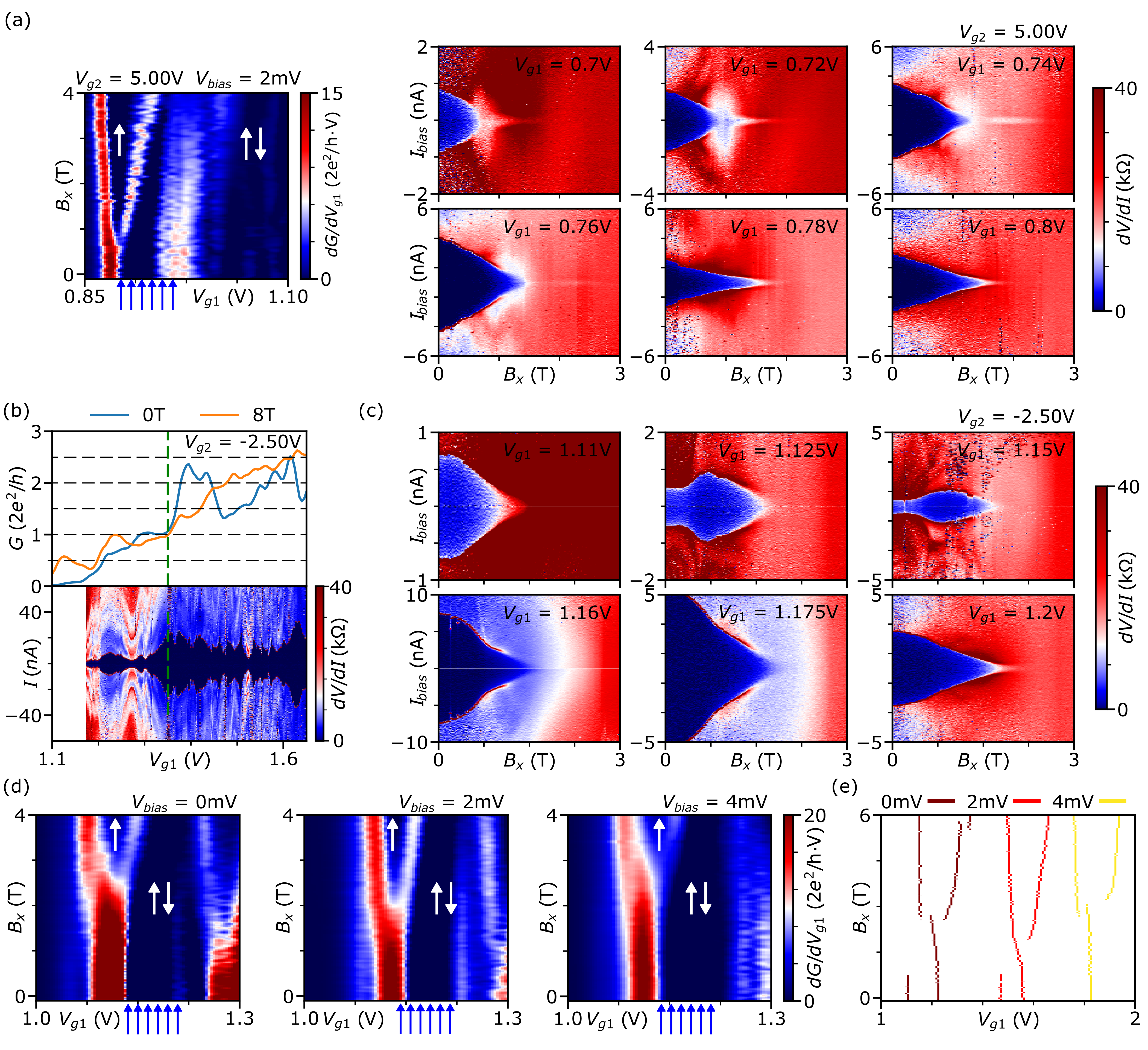}
\caption{\label{figs_QPC5_Ic_1st} (a) Differential resistance $R$ =  $dV/dI$ as a function of current bias and external parallel field $B_x$. The contact gate voltage $V_{g2}$ is fixed at 5~$\mathrm{V}$ to have totally six full transverse mode occupied in the contact region. The junction gate $V_{g1}$ varies for each panel but are all in the voltage range corresponding to the first mode. A transconductance plot is used to illustrate where the diffraction pattern is taken with blue arrows, with the indicated $V_{g1}$ all shifted +0.2~$\mathrm{V}$ due to gate dopping over measurements. (b) Gate dependence of conductance and critical current plotted side by side to identify where the first mode supercurrent is. The conductance trace is extracted from Fig.~\ref{figs_QPC5_QPC} (d) and (e) by taking line cut at $V_{bias}$ = 2~$\mathrm{V}$ at $B_x$ = 0~$\mathrm{T}$ plot and $V_{bias}$ = 0~$\mathrm{V}$ at $B_x$ = 8~$\mathrm{T}$ plot. Similar to Fig.~\ref{Fig2_1st_Ic} in main text but the $V_{g2}$ = -2.5~$\mathrm{V}$. to have three full transverse mode occupied in the contact region. (c) Same as (a) but the contact gate voltage is fixed at $V_{g2}$ = -2.5~$\mathrm{V}$ (d) Transconductance as function of $B_x$ and junction gate voltage $V_{g1}$ at fixed $V_{g2}$ = -2.5~$\mathrm{V}$, taken at different voltage $V_{bias}$. Blue arrows are used to illustrated where diffraction pattern are taken. (e) Extracted boundary between spin-polarized and spin-degeneracy region from (d), plotted with offset equals to 0.3~$\mathrm{mV}$ for each $V_{bias}$.}
\end{figure}

In Fig.~\ref{figs_QPC5_Ic_map} (b), we present extended $I_{sw}$ maps at various lead gate voltages $V_{g2}$. The number of conduction channels, modulated by the junction gate $V_{g1}$, is marked in each plot. From this panel, we observe electron spin-polarization suppression across most $V_{g2}$ values, akin to Fig.~\ref{Fig3_triplet_Ic}. For those not exhibiting suppression, we attribute this to the range of $V_{g1}$ being insufficient to encompass the entire first mode supercurrent. By comparing these plots, we note that the decay rate of $I_{sw}$ is relatively slower when $V_{g2}$ is smaller, aligning with the simulation results in Fig.~\ref{Fig5_Ic_sim} in the main text. These simulations demonstrate that a larger chemical potential in the lead region can induce a faster decay of $I_{sw}$ when the junction mode is fixed at one.

In Fig.~\ref{figs_QPC5_Ic_map} (c), we studied lead gate voltage effect by varying $V_{g1}$ and $V_{g2}$ together, keeping the junction in the first mode, to observe how different lead gate voltages affect the transport of the Josephson current. We conducted three scans following three dashed line traces, all maintaining a global electron mode equal to one. The panels in the yellow and pink boxes align with our interpretation that a higher gate voltage results in stronger suppression of $I_{sw}$ in the presence of an external field. However, the panel in the green box did not exhibit such results. As per Fig.~\ref{figs_QPC5_QPC_gvsg} (a), we observe that the conductance region corresponding to the first mode is not smooth and flat. This can be attributed to the nanowires not being purely ballistic, and the potential for a quantum dot to be driven by the gate, thereby inducing varying scatterings. Consequently, the $I_{sw}$ map might reflect superconducting behavior in a quantum dot regime rather than the first mode regime.

\subsection{Diffraction pattern of first mode supercurrent}
In Fig.\ref{figs_QPC5_Ic_1st}, we present the diffraction pattern of the first mode supercurrent current in a parallel magnetic field. Fig.~\ref{figs_QPC5_Ic_1st}(a) shows measurements taken at $V_{g2}$ = 5~$\mathrm{V}$ across various $V_{g1}$ values. On the right side of panel (a), we have displayed the transconductance again (adapted from Fig.~\ref{Fig3_triplet_Ic} and Fig.~\ref{figs_QPC5_QPC}) to highlight where the diffraction pattern is taken. The local bottom gate $V_{g1}$ shifts over time, which we attribute to the doping of the dielectric layer during the measurements. At the time the diffraction pattern was taken, the gate voltage corresponding to the first mode supercurrent ranged from 0.7 to 0.9~$\mathrm{V}$, whereas during the conductance plot acquisition, it ranged from 0.9 to 1.1~$\mathrm{V}$. Therefore, we conclude the gate shifts approximately 0.2~$\mathrm{V}$ during the measurements.

Comparing the boundaries between the spin-polarized and spin-degenerate regions in the transconductance plot and the diffraction pattern, we observe that although the decay rate of $I_{sw}$ doesn't have a significant difference, the superconductivity does disappear more rapidly at relatively smaller gate voltages. This can be attributed to the suppression from the polarization of electron, which is better elucidated in Fig.~\ref{Fig3_triplet_Ic}. We note a dip in the differential resistance beyond the critical field at $V_{g1}$ = 0.74~$\mathrm{V}$, which we currently do not have a clear interpretation for.

Fig.~\ref{figs_QPC5_Ic_1st}(b) is similar to Fig.~\ref{Fig2_1st_Ic} in the main text but at a different gate voltage, $V_{g2}$ = -2.5~$\mathrm{V}$. We concurrently plot the gate dependence of the differential conductance and the switching current, using the magnitude of $G$ as a guide to pinpoint the gate range where the supercurrent flows through a single conduction channel. In Fig.~\ref{figs_QPC5_Ic_1st}(c), we display several diffraction patterns of $I_{sw}$ at various $V_{g1}$ values at a fixed $V_{g2}$ = -2.5~$\mathrm{V}$. These diffraction patterns were measured immediately after the conductance scan, so no shift of gate voltage was observed.

We measured the transconductance as a function of $B_x$ and $V_{g1}$ at $V_{g2}$ = -2.5~$\mathrm{V}$ at several different voltage biases $V_{bias}$. We acknowledge that the Zeeman scan cannot provide an accurate boundary between the spin-polarized and degenerate regimes as the charging energy from $V_{bias}$ also plays a role. Nevertheless, we found that the scans we took were not comprehensive enough to rule out the effect from $V_{bias}$ and to accurately define the voltage range where the supercurrent reaches the spin-triplet regime.
\clearpage

\section{Measurement results in Device B}
\subsection{Gate dependence}
\begin{figure}[H]
\includegraphics[width = \textwidth]{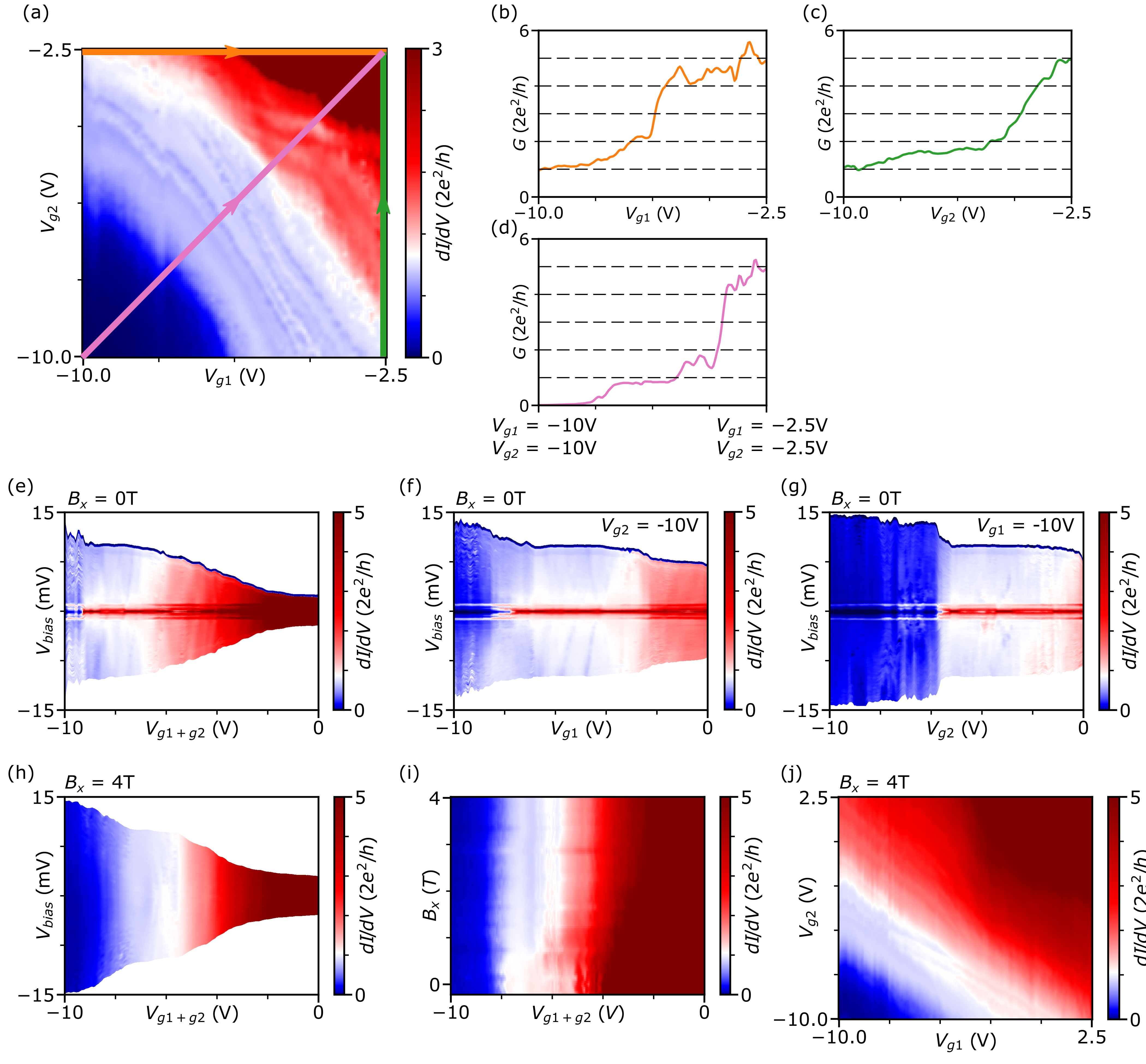}
\caption{\label{figs_QPC3_gate} (a) Differential conductance $G$ = $dI/dV$ as a function of junction gate $V_{g1}$ and contact gate $V_{g2}$, measured with fixed voltage bias $V_{bias}$ = 2mV and external field $B_x$ = 0T. (b)-(d) Three line cuts taken from (a). The color of plotted line is corresponding the linecut direction in (a). (e) Differential conductance $G$ = $dI/dV$ as a function of $V_{g1}$ + $V_{g2}$ (Varying both gates with same voltage) and $V_{bias}$. (f) $G$ as a function of $V_{g1}$ and $V_{bias}$ at fixed $V_{g2}$ = -10~$\mathrm{V}$. (g) $G$ as a function of $V_{g2}$ and $V_{bias}$ at fixed $V_{g1}$ = -10~$\mathrm{V}$. A series resistance $R_{in}$ = 7.1 k$\Omega$ is subtracted from $Vbias$ in the plot. (h) Differential conductance $G$ = $dI/dV$ as a function of $V_{g1}$ + $V_{g2}$ and $V_{bias}$ at fixed parallel field $B_x$ = 4~$\mathrm{T}$. (i) $G$ as function of parallel field $B_x$ and $V_{g1}$ + $V_{g2}$, measured at $V_{bias}$ = 2~$\mathrm{mV}$. (j)$G$ as a function of $V_{g1}$ and $V_{g2}$, measured at $V_{bias}$ = 2mV and $B_x$ = 4T.}
\end{figure}

Device B, is fabricated on chip QPC3, exhibits both junction gate $V_{g1}$ and lead gate $V_{g2}$ as effective controls of the conduction channels in the nanowires (Fig.~\ref{figs_QPC3_gate} (a) - (d)). However, both gates need to be lower than the safe voltage range (-5~$\mathrm{V}$ to 5~$\mathrm{V}$) to fully pinch-off the junction, potentially leading to over-doping of the dielectric layer and leakage between the gate and the junction. Consequently, Device B's gate continually shifted during the measurements, making it difficult to pinpoint a region corresponding to a single or few-mode regime. Hence, this device is not using for studying electron transport in specific electron modes.

In Fig.~\ref{figs_QPC3_gate}(e) to (g), we present 2D scans of the differential conductance $G$ at zero external field as a function of both gates, only $V_{g1}$, and only $V_{g2}$. We observe that the gate voltage region corresponding to the first mode while varying both local bottom gates $V_{g1+g2}$ is shifted compared to panels (a) and (d). However, the conductance region remains quantized in units of 1$G_0$. In panel (h), we scan conductance as a function of $V_{bias}$ and $V_{g1 +g2}$ at a parallel field $B_x$ = 4~$\mathrm{T}$, but we do not observe a clear diamond-shaped conductance plateau with a magnitude of 0.5$G_0$. In panel (i), we fail to detect the Zeeman split and emergence of a half-integer plateau corresponding to the spin-polarized regime. In panel (j), we note a slight shift in gate dependence of conductance from panel (a), but no evidence of a spin-polarized state is found.

\subsection{Diffraction pattern of switching current}
\begin{figure}[H]
\includegraphics[width = \textwidth]{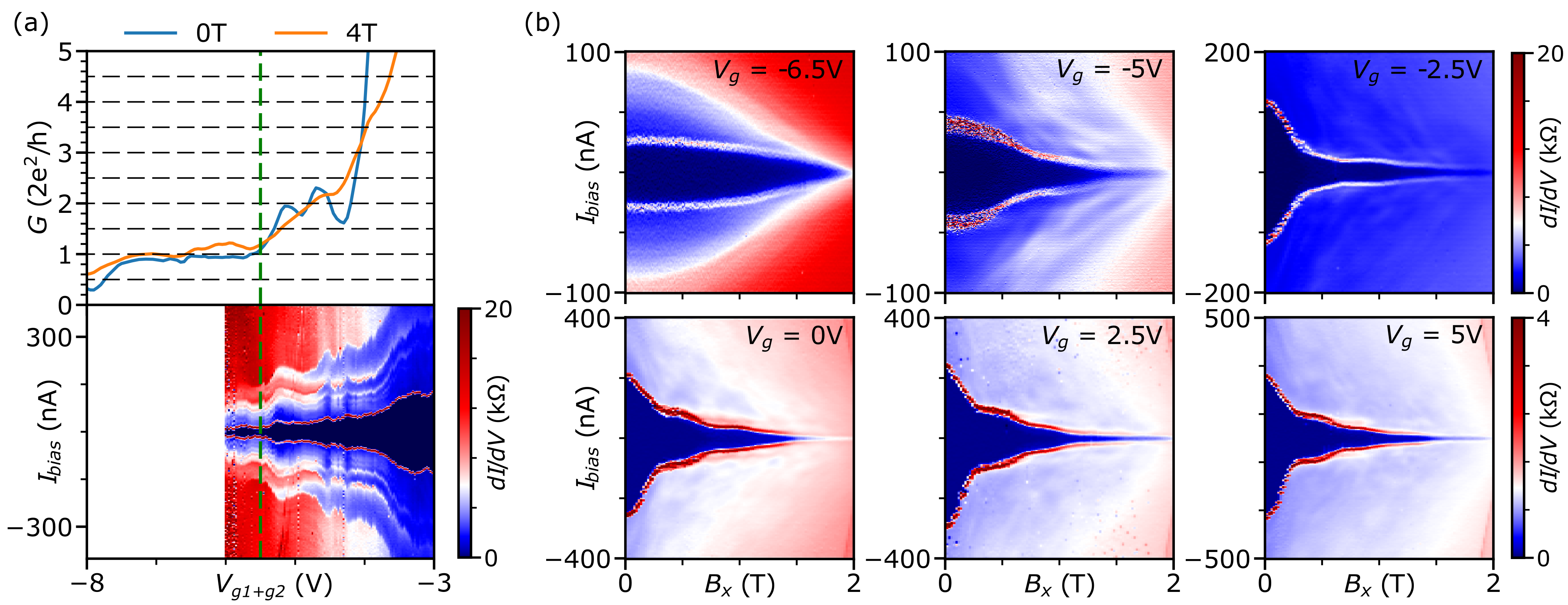}
\caption{\label{figs_QPC3_Ic}  (a) Gate dependence of conductance and switching current plotted side by side. The conductance trace is extracted from Fig.~\ref{figs_QPC3_gate} (e) and (h) by taking the line cut at $V_{bias}$ = 2~$\mathrm{mV}$ at $B_x$ = 0~$\mathrm{T}$ and $V_{bias}$ = 0~$\mathrm{mV}$ at $B_x$ = 4~$\mathrm{T}$. The boundary between first mode and higher modes is labeled with green dashed line. (b) Diffraction pattern of supercurrent in parallel field $B_x$ at different fixed gate voltage $V_g$ ($V_{g1 + g2}$).}
\end{figure}

In Fig.~\ref{figs_QPC3_Ic} (a), we present side-by-side plots of conductance and switching current $I_{sw}$ as functions of gate voltage, which allow us to identify the gate voltage range where only the first mode supercurrent is transported through the junction. In panel (b), we display supercurrent diffraction patterns at various fixed gate voltages, where $V_g$ represents all gates tuned together to the same value, identical to $V_{g1+g2}$. When $V_{g}$ = -6.5$\mathrm{V}$, the junction has only one conduction channel and one spin-full transverse mode available, and the supercurrent decays more slowly compared to larger gate voltages. However, we observe that the critical field at which superconductivity disappears remains the same across all gate voltage values.
\clearpage

\section{Full measurement data of Device C}
\begin{figure}[H]
\includegraphics{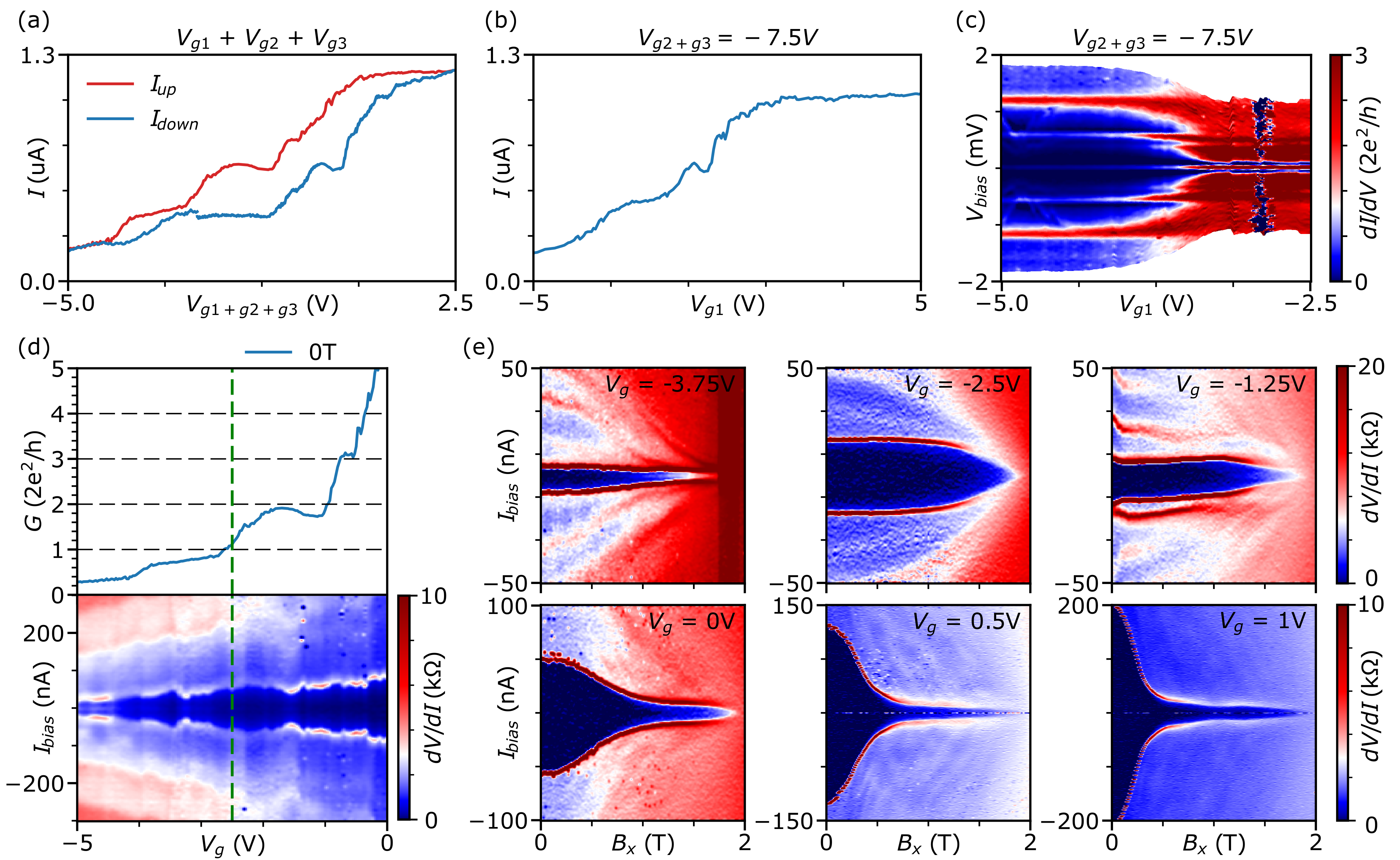}
\caption{\label{figs_QPC4_data} Full measurement Data in Device C. (a)-(b)Current through the device as a function of all gate voltage (a), $V_{g1}$ that is under the junction region (b). (c) $G$ as a function of $V_{bias}$ and $V_{g1}$ when other two gates $V_{g2}$ and $V_{g3}$ are set to be -7.5~$\mathrm{V}$. (d) Gate dependence of conductance and switching current are plotted side by side. The conductance trace is converted from the red $I$-$V_g$ trace in panel(a) with a sereis resistance $R_s$ = 7.1$\mathrm{k\Omega}$ is used in calculation. All three gate voltages are set to be the same value and labeled as $V_g$ in panels. (e) Differential resistance as function of $I_{bias}$ and $B_x$. }
\end{figure}
Similar to Device B, the tunnel junction in Device C cannot be fully closed by the local gates within the safe voltage range, and we observed more serious leakage when $V_g$ falls below -5~$\mathrm{V}$. Consequently, this device was not used for QPC studies, although we identified well-defined multiple Andreev states, indicating a lack of scattering in the system. As a result, we cannot confidently assert the exact number of modes at each gate voltage. However, estimations based on the pinch-off trace in Fig.~\ref{figs_QPC4_data} (a), combined with resistance measurements from the normal state in diffraction pattern plots, assist in determining the number of transverse modes in Fig.~\ref{Fig4_Ic_map}(c) in the main text.

In Fig.~\ref{figs_QPC4_data} (d), we attempted to identify the gate voltage region associated with the first mode supercurrent. In Fig.~\ref{figs_QPC4_data} (e), we present several diffraction patterns of $I_{sw}$ when a parallel field is applied. When $V_g$ = -2.5~$\mathrm{V}$, both conductance and normal state resistance indicate that only one transverse mode is available, and we observe a slower supercurrent compared to the higher gate voltage regime. These findings align with the $I_{sw}$ map in the main text (Fig.~\ref{Fig4_Ic_map}) and the phenomena we observed in Device B.
\clearpage

\section{Full measurement data of Device D}
\begin{figure}[H]
\includegraphics{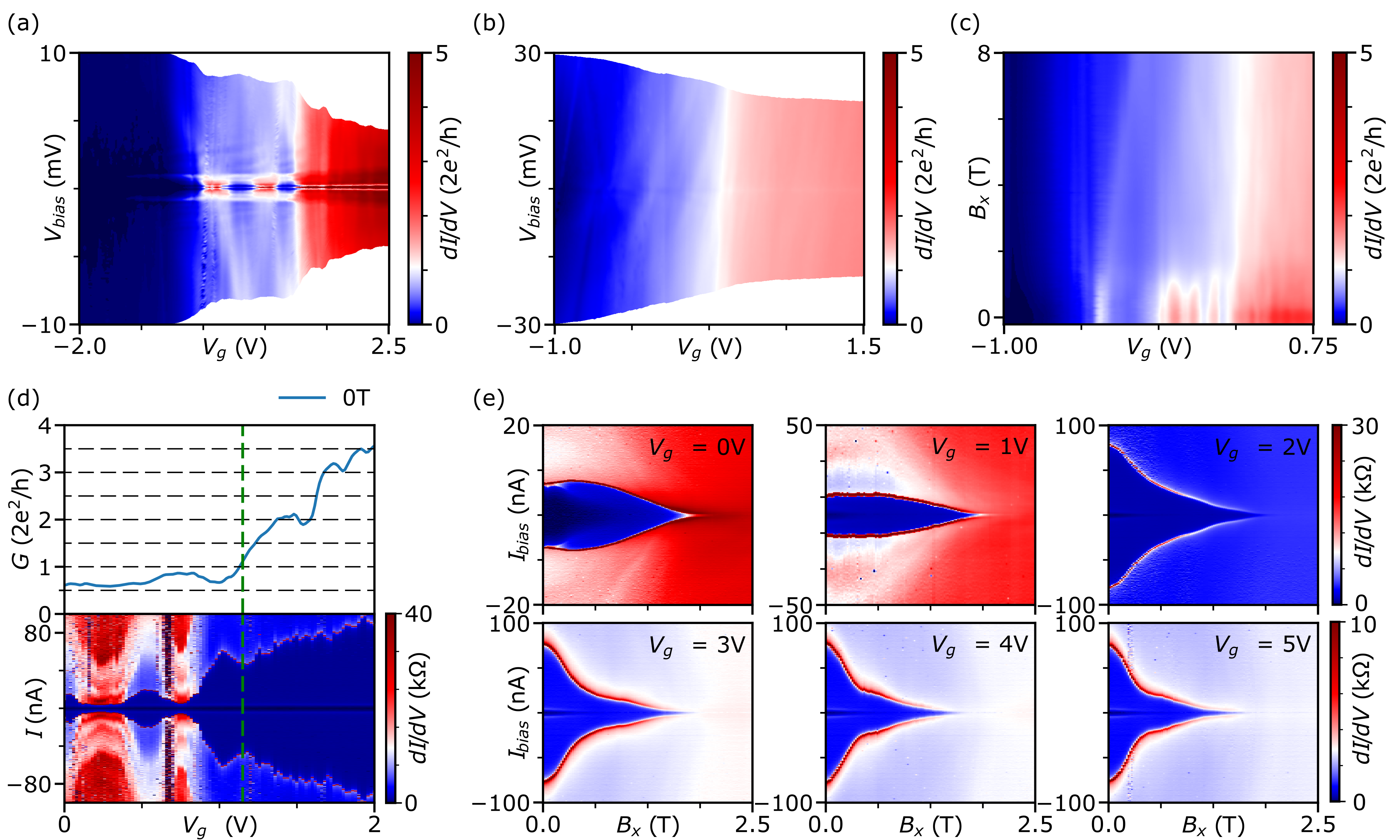}
\caption{\label{figs_deviceD_data} Full measurement Data in Device D. (a), (b) Differential conductance $G$ = $dI/dV$ as a function of $V_{g}$ and $V_{bias}$ at external field $B_x$ = 0~$\mathrm{T}$ and 8~$\mathrm{T}$. A series resistance $R_{in}$ = 4.1~$\mathrm{k\Omega}$ is used to calculate the conductance and subtracted from the $V_bias$. (c) $G$ as a function of of $V_{g}$ and $B_x$, Voltage bias is set to be $V_{bias}$ = 0~$\mathrm{mV}$. (d) Gate dependence of conductance and switching current are plotted side by side. The conductance trace is extracted from the panel(a) by taking linecut at $V_{bias}$ = 2~$\mathrm{mV}$. (e) Differential resistance as function of $I_{bias}$ and $B_x$. $V_{g2}$ Varies to different values.}
\end{figure}

Device D only has one gate that can adjust the chemical potential in the nanowire junction and it looks like a N-S-N island device under SEM (see Fig.~\ref{figs_devices}). Nevertheless, sharp transitions in the differential resistance 2D scan signify the presence of a switching current. At a parallel field $B_x$ = 0~$\mathrm{T}$, there is a conductance region with an almost quantized magnitude of 1$G_0$ for gate voltages ranging from 0 to 1~$\mathrm{V}$ (Fig.\ref{figs_deviceD_data} (a)). At a parallel field $B_x$ = 8$\mathrm{T}$, an additional conductance region appears with a magnitude of 0.5$G_0$(Fig.~\ref{figs_deviceD_data} (b)).

In Fig.~\ref{figs_deviceD_data} (c), we plot $G$ as a function of $V_g$ and $B_x$ and observe a transition between a spin-degeneracy-only state and a spin-polarized regime, indicating a Zeeman split as evidence of a QPC. However, we detect two switching current peaks across all gate voltages, with the first transition being non-adjustable. Given the device's geometry seen in the SEM image, which more closely resembles an island device than a Josephson junction device, we have treated data from this device as supplementary material rather than incorporating it into the main text.

In Fig.~\ref{figs_deviceD_data} (e), we present several diffraction patterns of the switching current at different gate voltages when a parallel field is applied. At gate voltages $V_{g1}$ = 0~$\mathrm{V}$ and 1~$\mathrm{V}$, both the conductance and the normal state resistance indicate the presence of only one available transverse mode in the nanowire junction. Interestingly, we observe a slower $I_{sw}$ decay compared to patterns taken at higher gate voltages.
\clearpage

\section{Full measurement data of Device E (NbTin-InSb naowires)}
\subsection{QPC in Device E}
\begin{figure}[H]
\includegraphics{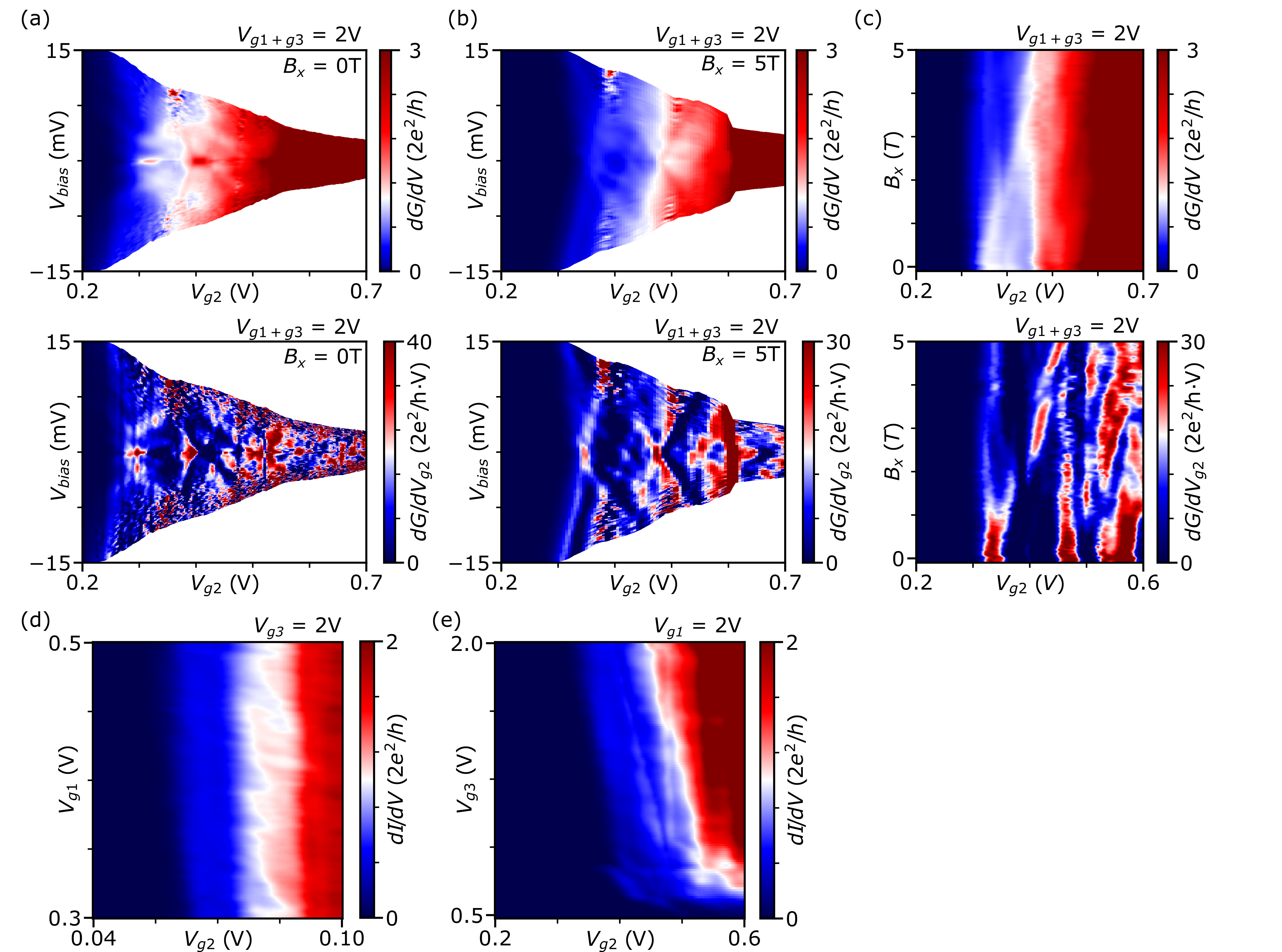}
\caption{\label{figs_SC2_gate_dependence} QPC and gate dependence in Device D. (a), (b)Upper: Differential conductance $G$ = $dI/dV$ as a function of $V_{g}$ and $V_{bias}$ at external field $B_x$ = 0~$\mathrm{T}$ and 5~$\mathrm{T}$, contact gate voltages are set to be $V_{g1}$ and $V_{g3}$ = 2V. A series resistance $R_{in}$ = 6.5k$\Omega$ is used to calculate the conductance and subtracted from the $V_bias$. Lower: Transconductance by taking $dG/dV_{g2}$ from upper plot for the boundary of quantized conductance region. (c) Upper: $G$ as a function of of $V_{g2}$ and $B_x$, Voltage bias is set to be $V_{bias}$ = 2~$\mathrm{mV}$, $V_{g1}$ and $V_{g3}$ = 2V. Lower: Transconductance by taking $dG/dV_{g2}$ from upper plot. (d) Left: Differential conductance $dI/dV$ as a function of junction gate $V_{g1}$ and contact gate $V_{g2}$, measured with fixed voltage bias $V_{bias}$ = 2mV, $V_{g3}$ = 2~$\mathrm{V}$ and external field $B_x$ = 4~$\mathrm{T}$. (e) $G$ as a function of junction gate $V_{g3}$ and contact gate $V_{g2}$, measured at $V_{g1}$ = 2~$\mathrm{V}$ and $B_x$ = 4~$\mathrm{T}$. }
\end{figure}

Devices E and F were fabricated by ex-situ depositing an NbTiN film onto InSb nanowires to form superconducting leads. The NbTiN shell is not as smooth as that of Tin, but the bulk NbTiN has a significantly higher critical temperature and field compared to Tin or Aluminum. Device E was fabricated on a local bottom gate chip, which is why we have better control of the junction and are able to observe the QPC in this device, unlike in Device F.

Fig.~\ref{figs_SC2_gate_dependence} provides evidence of a QPC in Device E. Panel (a), taken at $B_x$ = 0~$\mathrm{T}$, reveals a diamond-shaped conductance region with a uniform magnitude of 1$G_0$. Taking the derivative with respect to the gate voltage $V_{g2}$ clearly demonstrates the boundaries of this quantized conductance region. In panel (b), the scan conducted at $B_x$ = 5~$\mathrm{T}$ shows a diamond-shaped conductance region of roughly the same size as the plateau at 0~$\mathrm{T}$, but with a conductance magnitude of 0.5$G_0$, which indicates this is corresponding to the spin-polarized regime

In panel (c), we hold $V_{bias}$ constant at 2~$\mathrm{mV}$ and plot $G$, observing the transition between the spin-degeneracy and spin-polarized regimes. Panels (d) and (e) explore the effects of the lead gate voltages $V_{g1}$ and $V_{g3}$. We find that both contribute weaker effects compared to the junction gate voltage $V_{g2}$. Therefore, to avoid scattering from the quantum dot, they are both set to their highest positive value of 2~$\mathrm{V}$.

\subsection{supercurrent in Device E}
\begin{figure}[H]
\includegraphics{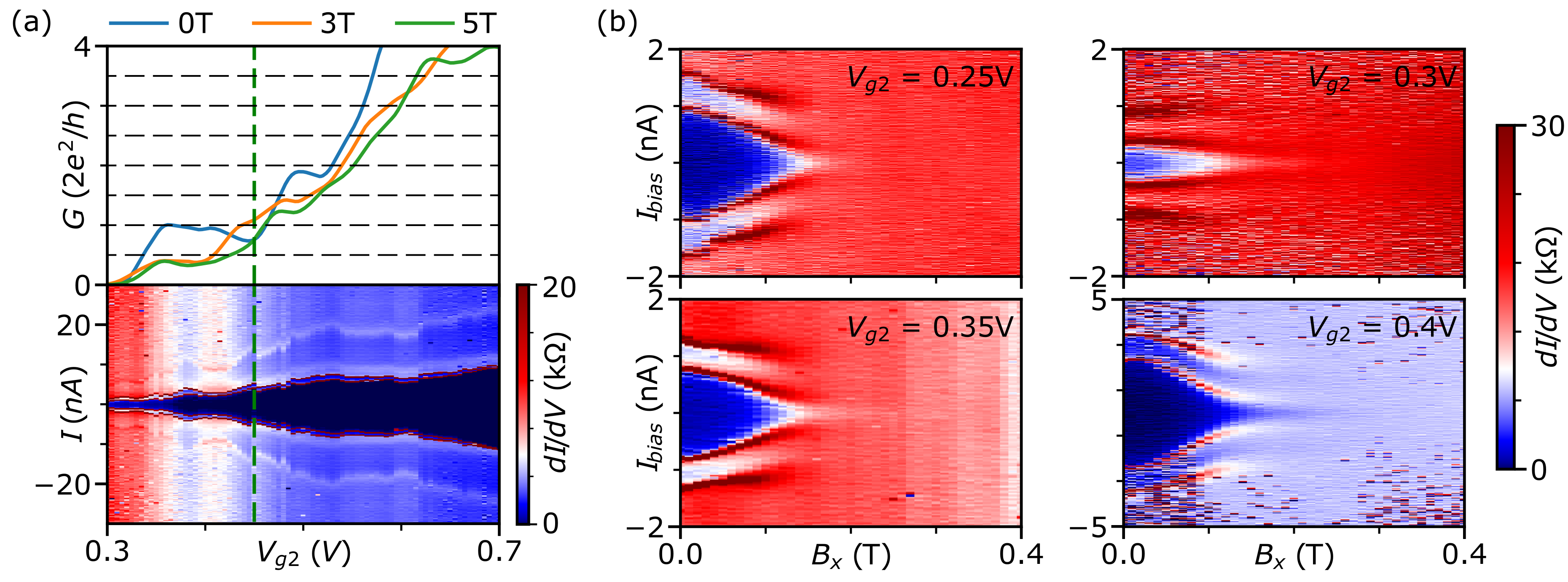}
\caption{\label{figs_SC2_ic} (a) Differential conductance $G (dI/dV)$ taken at normal state as functions of $V_{g2}$ at 0~$\mathrm{T}$/3~$\mathrm{T}$/5~$\mathrm{T}$ and gate sweep of $I_{sw}$ at 0T are plotted next to each other, extracted from Fig.~\ref{figs_SC2_gate_dependence} (c) by taking line cuts. Gate scan of supercurrent is plotted side by side.Green dashed line is used to indicate the boundary between first mode and more modes. (b) Differential resistance as function of $I_{bias}$ and $B_x$. $V_{g2}$ Varies to different values but in the voltage range that corresponding to first electron mode. Contact gates voltage are fixed to $V_{g1}$ and $V_{g3}$ = 2~$\mathrm{V}$.}
\end{figure}
Fig.\ref{figs_SC2_ic} presents both gate and field scans of supercurrent in Device E. In panel (a) we study the gate range for first mode supercurrent. Here, we find that a gate voltage $V_{g2}$ from 0.3~$\mathrm{V}$ to 0.45~$\mathrm{V}$ corresponds to the presence of only one conduction channel in the nanowire device with supercurrent transport through the junction.

In panel (b), we study diffraction patterns of supercurrent at several fixed gate voltages within the first mode range. We observe that although the supercurrent decays more slowly, it has a much smaller critical field compared to the nanowire junctions in Ref.~\cite{zuo2017supercurrent} and Device F.
\clearpage

\section{Full measurement data of Device F (NbTin-InSb naowires)}
\begin{figure}[H]
\includegraphics{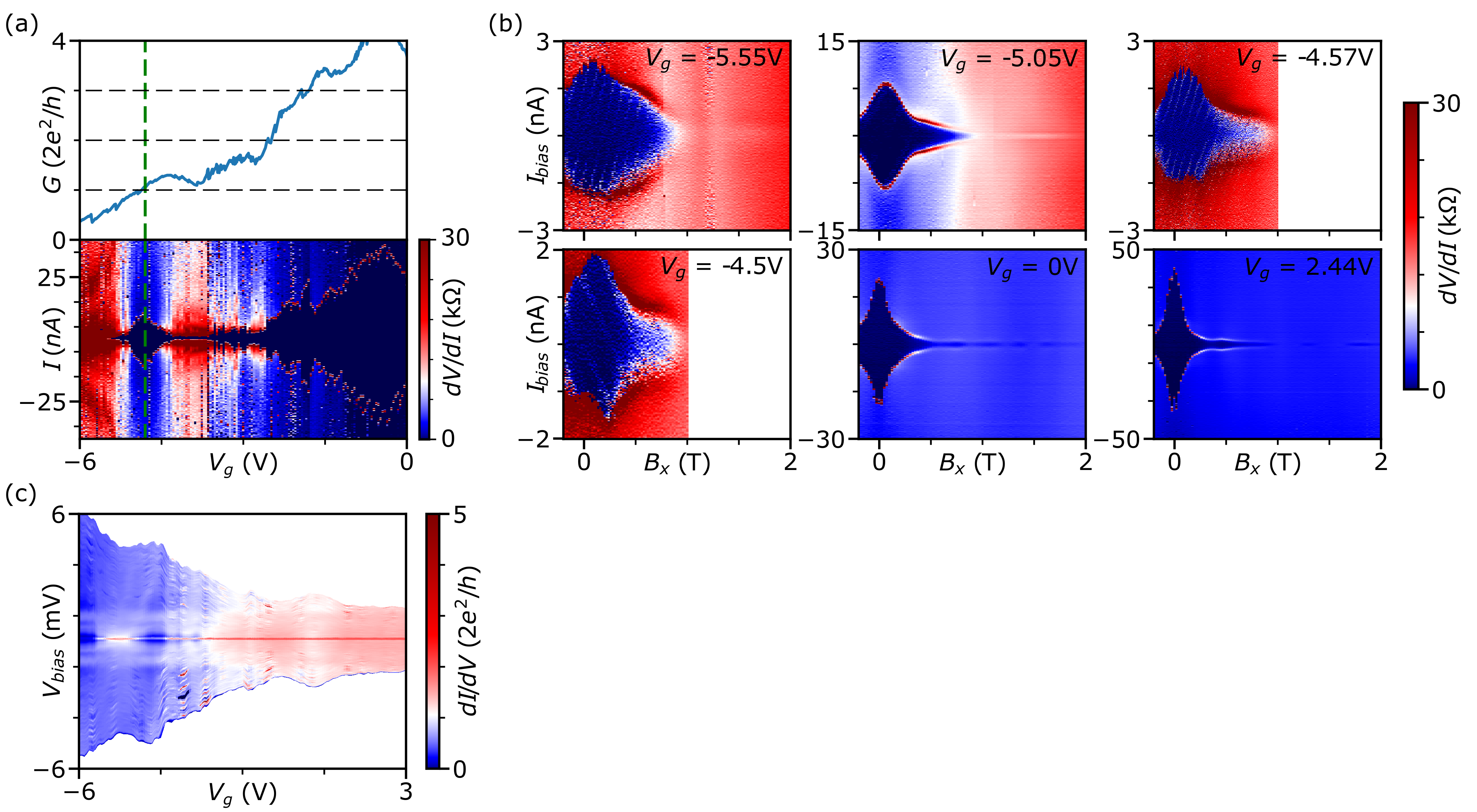}
\caption{\label{figs_Hao_data} Full measurement Data in Device F. (a) Current through the device as a function of global gate voltage $V_{g}$. (b) Differential conductance $G$ = $dI/dV$ as a function of $V_{g}$ and $V_{bias}$ at external field $B_x$ = 0~$\mathrm{T}$. A series resistance $R_{in}$ = 13.9k$\Omega$ is used to calculate the conductance and subtracted from the $V_{bias}$. (c) Differential resistance as function of $V_{g}$ and $I_{bias}$ for the gate sweep of $I_{sw}$ at 0T. (d) Differential resistance as function of $I_{bias}$ and $B_x$. $V_{g}$ Varies to different values in each panel.}
\end{figure}

Device F is the only device fabricated with a bottom global back-gate chip (see Fig.~\ref{figs_devices} (f)). The chemical potential across all regions of the devices is tuned simultaneously. As there is no conductance plateau quantized to 1$G_0$ at zero parallel field (Fig.~\ref{figs_Hao_data} (a) and (c)), we do not have a quantum point contact (QPC) for precise control over the number of electron transverse modes in the nanowires.

In panel (a), we identify the gate voltage range where both resistance and conductance indicate the presence of only one conduction channel in the nanowires and there is supercurrent within this range. Upon setting the gate voltage $V_g$ to a value in this range, we observe that the supercurrent decays more slowly compared to the diffraction pattern measured at higher gate voltages. Additionally, an irregular disappearance and reappearance of supercurrent is observed at $V_{g}$ = 0~$\mathrm{V}$ and 2.44~$\mathrm{V}$, which is similar to the findings reported in Ref.~\cite{zuo2017supercurrent}.
\clearpage

\section{$I_{sw}R_n$ product comparison between all devices}
\begin{figure}[H]
\includegraphics{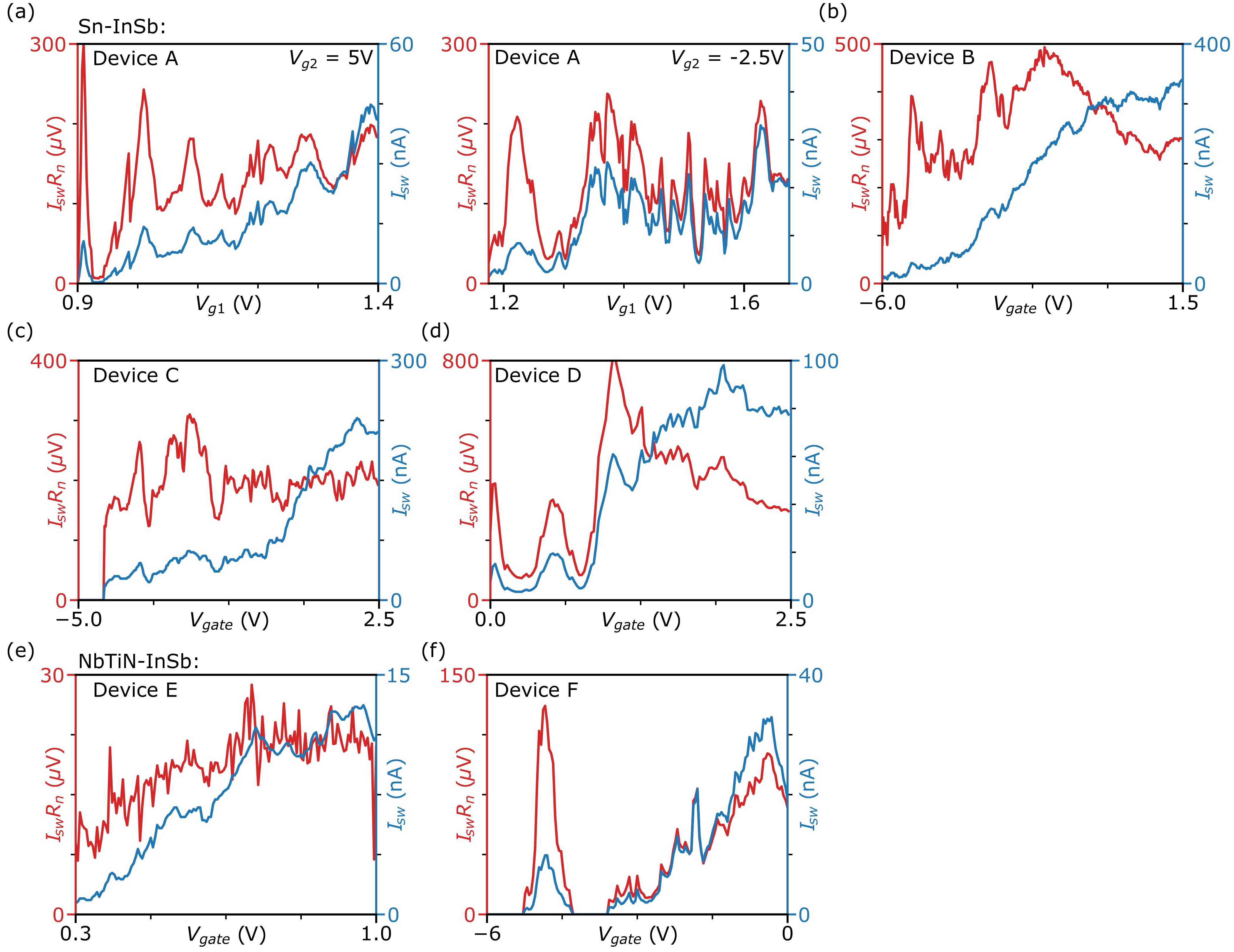}
\caption{\label{figs_IcRn} Critical current Normal state resistance product $I_{sw}R_N$ as a function of gate voltage measured in each devices. (a)-(f) Device A - F.}
\end{figure}

The strength of the Josephson effect in the nanowire junction, induced by the proximity effect, is quantified using the product of the switching current and normal state resistance, $I_{sw}R_n$, as shown in Fig.~\ref{figs_IcRn}. In panel (a), we present the measurement results at two different lead gate voltages, $V_{g2}$ = 5~$\mathrm{V}$ and -2.5~$\mathrm{V}$. The normal state resistance is taken from the QPC scans in Fig.~\ref{figs_QPC5_QPC}. The magnitude of the switching current $I_{sw}$ is extracted from Fig.~\ref{Fig2_1st_Ic} (5~$\mathrm{V}$) and Fig.~\ref{figs_QPC5_Ic_1st}(b) (-2.5~$\mathrm{V}$) by tracking from the zero current bias and identifying the first point where the differential resistance exceeds 2~$\mathrm{k\Omega}$ at positive and negative bias for each gate voltage. The plot reveals that at both lead gate voltages, the $I_{sw}R_N$ product is smoothly vibrating around 200~$\mathrm{\mu V}$, comparable to the energy gap of Tin ($\Delta$ $\approx$ 650~$\mathrm{\mu eV}$).

The normal state resistance in panels (b), (c), and (d) is obtained from Fig.~\ref{figs_QPC3_gate} (d), Fig.~\ref{figs_QPC4_data} (d), and Fig.~\ref{figs_deviceD_data} (a). The magnitude of supercurrent is read from Fig.~\ref{figs_QPC3_Ic}(a), Fig.~\ref{figs_QPC4_data}(d), and Fig.~\ref{figs_deviceD_data}(d), respectively. Their $I_{sw}R_n$ products are larger than Device A, indicating a stronger Josephson effect in the junction, and all comparable to the energy gap of Tin.

Conversely, Devices E and F, fabricated with NbTiN and InSb nanowires, show a relatively smaller $I_{sw}R_n$ product and weaker Josephson effect in panels (e) and (f), despite NbTiN having a larger superconducting gap and higher critical temperature. In panel (e), we derive the normal state resistance from Fig.~\ref{figs_SC2_gate_dependence} (a) and the magnitude of $I_{sw}$ from Fig.~\ref{figs_SC2_ic}(a), the $I_{sw}R_n$ product approximates 20~$\mathrm{\mu V}$. In panel (f), we extract both the $R_n$ and $I_{sw}$ from the conductance trace in Fig.~\ref{figs_Hao_data} (a), the conductance trace is based on an $I$-$V_g$ trace measured at a fixed $V_{bias}$ = 10~$\mathrm{mV}$ and zero field. There is a high peak of $I_{sw}R_n$ around $V_{g}$ = -5~$\mathrm{V}$, and the $I_{sw}R_n$ product is not stable, varying with the gate voltage. We believe this instability is due to the gate being unstable, shifting between the $I$-$V_g$ measurement and the gate scan of $I_{sw}$. Nonetheless, the maximum magnitude of $I_{sw}R_n$ does not grow significantly, remaining considerably smaller than the energy gap of NbTiN ($\Delta >$ 3~$\mathrm{meV}$).

In conclusion, upon examining the $I_{sw}R_n$ products, we find that the superconductivity induced by the thin Sn shell in InSb nanowires is much stronger than that in NbTiN-InSb devices. As we've stated in the main text, stronger induced superconductivity is crucial for achieving topological states, as it must counteract the suppression from the Zeeman effect and orbital effect in the presence of a field. In our case, Sn-InSb nanowires present a more robust and consistent magnitude of $I_{sw}R_n$ across the gate voltage range. Furthermore, the Josephson current is well adjusted by the gate, making it a superior platform for studying Majorana-related Josephson effects.
\clearpage

\section{Simulation model}\label{sec:simulation}
\subsection{simulation model}
\begin{figure}[H]
\includegraphics{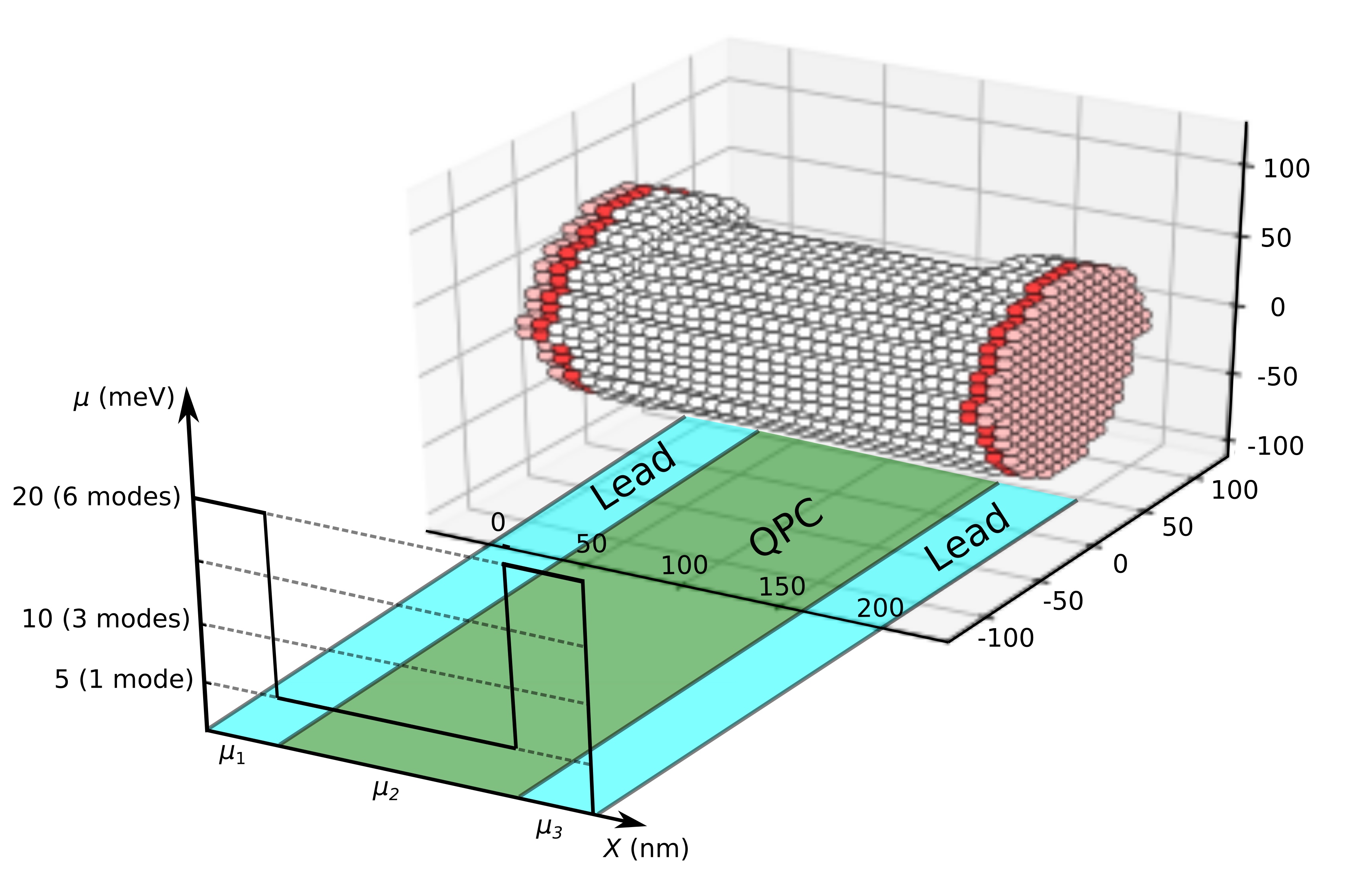}
\caption{\label{figs_model_geometry}. Tight-binding model generated by the simulation. Red dots represent the infinite leads. Lattice constant is 8~$\mathrm{nm}$. Shell is covering the nanowires to make contact with length of 24~$\mathrm{nm}$ at each end. Uncovered nanowires is 160~$\mathrm{nm}$ long. Three chemical potentials are defined locally at leads and junction regions. In this figure we are presenting the condition where we set $\mu_{leads}$ = 20~$\mathrm{meV}$ for six spin-full transverse mode and $\mu_{QPC}$ = 5~$\mathrm{meV}$ for one transverse mode.}
\end{figure}

To numerically simulate the superconductor-semiconductor-superconductor nanowires Josephson junction in the presence of external parallel magnetic field, we consider the following Hamiltonian for a nanowire that is covered by superconductor lead at both ends:

\begin{equation}
    H = \left(\frac{\mathbf{p}^2}{2m^*}-\mu(x) + \delta U\right)\tau_z +\alpha(p_x \sigma_y - p_y\sigma_x)\tau_z + g\mu_B\mathbf{B}\cdot \hat{\sigma} +\Delta \tau_x,\label{eq:hamiltonian}    
\end{equation}

where $\tau_i$ and $\sigma_i$ are Pauli matrices act on particle-hole and spin space, respectively. $\mathbf{p} = -i\hbar \nabla +e \mathbf{A}\tau_z$ is the canonical momentum, and the magnetic potential $\mathbf{A}$ is chosen to be $[B_yz-B_zy, 0, B_{x}y]^T$, so that it is invariant along the $x-$direction. Further, $m^*$ is the effective mass, $\mu$ is the chemical potential and $ \delta U$ represent the onsite disorder inside the nanowires. The Zeeman effect is given by $g\mu_B\mathbf{B}\cdot \hat{\sigma}$ and the Rashba spin-orbit coupling is given by $\alpha(p_x\sigma_y - p_y\sigma_x)$. Finally, $\Delta$ is the superconducting pairing potential. 

Chemical potential is defined locally, where we have:
\begin{equation}
  \mu(x) =
    \begin{cases}
      \mu_1 & \text{if 0 $<$ x $<$ $l_{leads}$}\\
      \mu_2 & \text{if $l_{leads}$ $<$ x $<$ $l_{leads}$+$l_{junction}$}\\
      \mu_3 & \text{if x $>$ $l_{leads}$+$l_{junction}$}
    \end{cases}       
\end{equation}

The geometry of the simulation model is plotted in Fig.~\ref{figs_model_geometry}. In the simulation, we give same value to the $\mu_1$ and $\mu_3$ and call them lead chemical potential, $\mu_2$ is the chemical potential of QPC. As this is a microscopic model, it facilitates the computation of electron hopping between neighboring sites. The process of deriving the critical current from the Hamiltonian is discussed in Ref.\cite{zhang2022evidence}. 
\clearpage

\subsection{Parameter dependence of QPC}
\begin{figure}[H]
\includegraphics{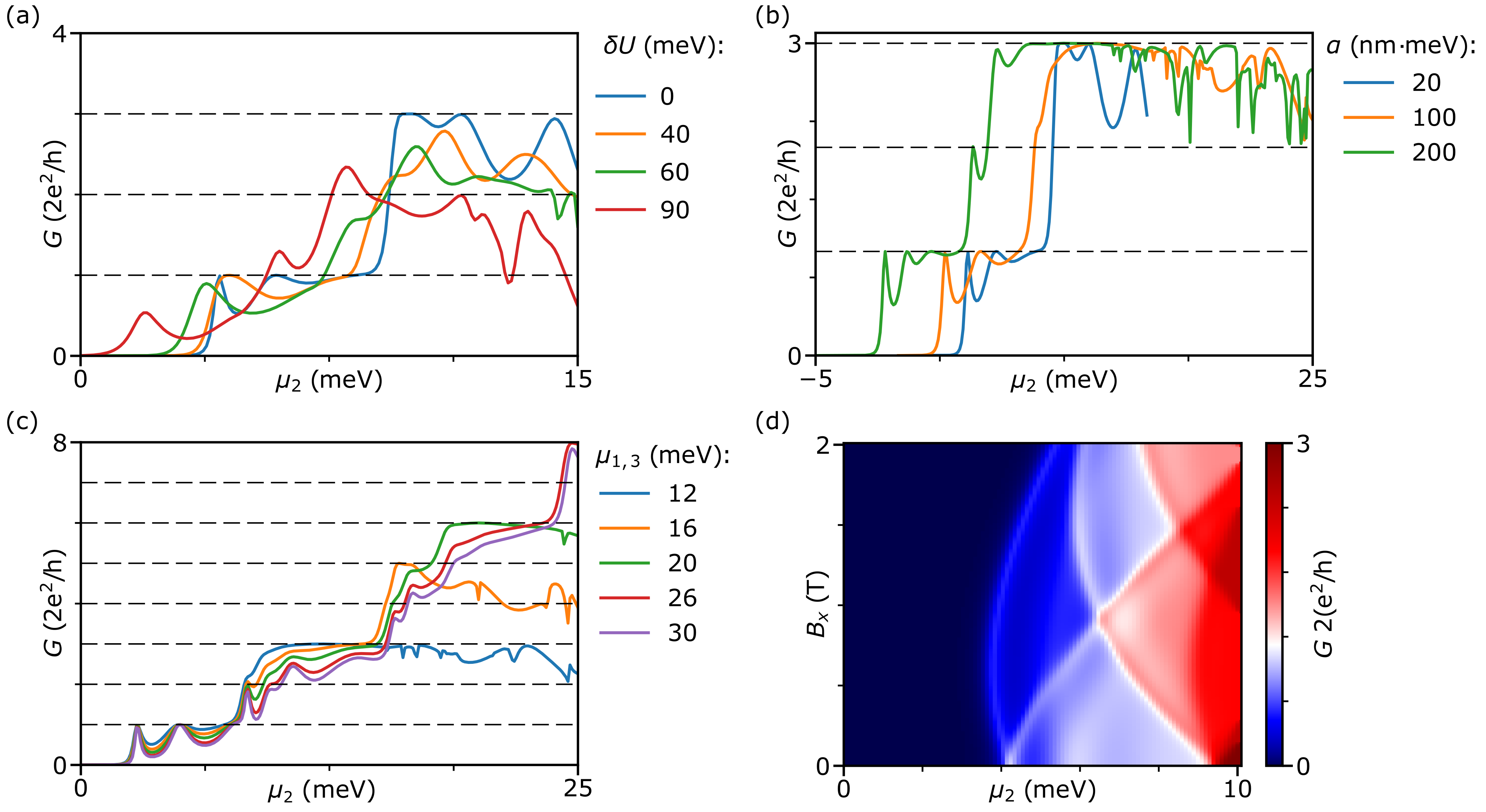}
\caption{\label{figs_sim_qpc}. Parameters dependence of constructing QPC in the simulation. (a) Disorder strength ($\delta U$) dependence. (b) Spin-orbit effect Strength ($\alpha$) dependecne. (c) Lead chemical potential $\mu_{lead}$ dependence. In panel (a)-(c), if not studied for dependence, the parameters are set to be: $\mu_{1,3}$ = 10~$\mathrm{meV}$, $\alpha$ = 100~$\mathrm{nm\cdot meV}$, temperature $T$ = 0.1~$\mathrm{T}$, disorder $\delta U $= 0~$\mathrm{meV}$. (d) Full conductance scan as function of $\mu_2$ and parallel field $B_x$ presented in Fig.~\ref{Fig3_triplet_Ic} (c) in the main text.}
\end{figure}

The parameters used in the simulation are analyzed from two perspectives: their impact on the construction of the Quantum Point Contact (QPC) and their influence on the supercurrent field evolution. This section primarily discusses the QPC dependency.

In Fig.~\ref{figs_sim_qpc}(a), we investigate the effect of disorder on the QPC conductance dependence, while keeping all other parameters constant.

We focused on the first plateau, which corresponds to a single spin-full transverse mode with a magnitude of 2~$\mathrm{e^2/h}$ or 1$G_0$. Our findings suggest that quantized first plateaus (in the unit of $G_0$) are only achievable when the disorder strength $\delta U$ is less than 40~$\mathrm{meV}$. An increase in disorder reduces the transmission rate between two leads, subsequently lowering the magnitude of conductance $G$ and introducing additional plateaus.

It's worth noting that our simulation lacks the plateaus of 2$G_0$ and 4,5$G-0$, potentially due to the subband's degeneracy at zero field, as discussed in Ref.~\cite{kammhuber2016conductance}. In our simulation, disorder is introduced by randomly mapping local defects into the junction, causing scattering during electron transportation. Ref.~\cite{zuo2017supercurrent} determined the disorder parameter by calculating the corresponding mean free path $l_{mfp}$ for different $\delta U$ strengths. They selected $\delta U$ = 90~$\mathrm{meV}$, yielding $l_{mfp}$ = 250~$\mathrm{nm}$, which aligns closely with the mobility and mean free path characterized in InSb nanowires~\cite{gul2015towards}. However, the QPC is highly influenced by disorder distribution. In our case, it suggests a junction that is nearly disorder-free in the junction region. Therefore, we selected $\delta U$ = 0~$\mathrm{meV}$ for the spin-polarization studies in Fig.~\ref{Fig3_triplet_Ic} in the main text to match the interpretation about 1st mode supercurrent in Ref.~\cite{zuo2017supercurrent}, and $\delta U$ = 40~$\mathrm{meV}$ for discussion in Fig.~\ref{figs_sim_ic}, which satisfies both the real-world condition of quasi-ballistic nanowires and the simulation condition of having a quantized plateau.

Fig.\ref{figs_sim_qpc}(b) discusses the effect of spin-orbit interaction strength on the QPC. We observe small difference in the conductance trace shape as a function of $\mu_2$, but it shifts towards a more negative chemical potential side as $\alpha$ increases. In Ref.~\cite{zhang2022evidence}, we studied how $\alpha$ impacts the supercurrent diffraction pattern and determined that the magnitude of approximately 100~$\mathrm{nm\cdot meV}$ is necessary to produce the skewed shape. As this research shares the same wires and chips as the current project, we conclude that $\alpha$ = 100~$\mathrm{nm\cdot meV}$ should be continuously used for the project's simulation.

In Fig.\ref{figs_sim_qpc}(c), we explore the influence of the lead chemical potential $\mu_{1,3}$ on the QPC and the number of transverse modes in the junction. When $\mu_{1,3}$ varies, the conductance trace is not shifted with respect to the QPC chemical potential $\mu_2$, but it only affects the maximum mode occupied in the Josephson Junctions (JJs). We employ the same method to identify how the chemical potential controls the available modes in the junction and conclude that $\mu$ = 5, 10, 20$\mathrm{meV}$ correspond to 1, 2, 3 occupied modes, respectively.

Fig.~\ref{figs_sim_qpc}(d) presents the Zeeman scan results from the simulation, incorporating parameters from the previous discussions. In this plot, we have a disorder of $\delta U$ = 0~$\mathrm{meV}$, a spin-orbit interaction strength of $\alpha$ = 100~$\mathrm{nm\cdot meV}$, and a lead chemical potential $\mu_{1,3}$ = 20~$\mathrm{meV}$. We observe a quantized plateau of 1$G_0$ at zero field, which splits into two plateaus of 0.5$G_0$ and 1$G_0$ due to the Zeeman effect. By taking the derivative with respect to the chemical potential $\mu_2$, we obtain the transconductance plot presented in the main text, Fig.~\ref{Fig3_triplet_Ic}(c).

\subsection{Parameter dependence of $I_c$}
\begin{figure}[H]
\includegraphics{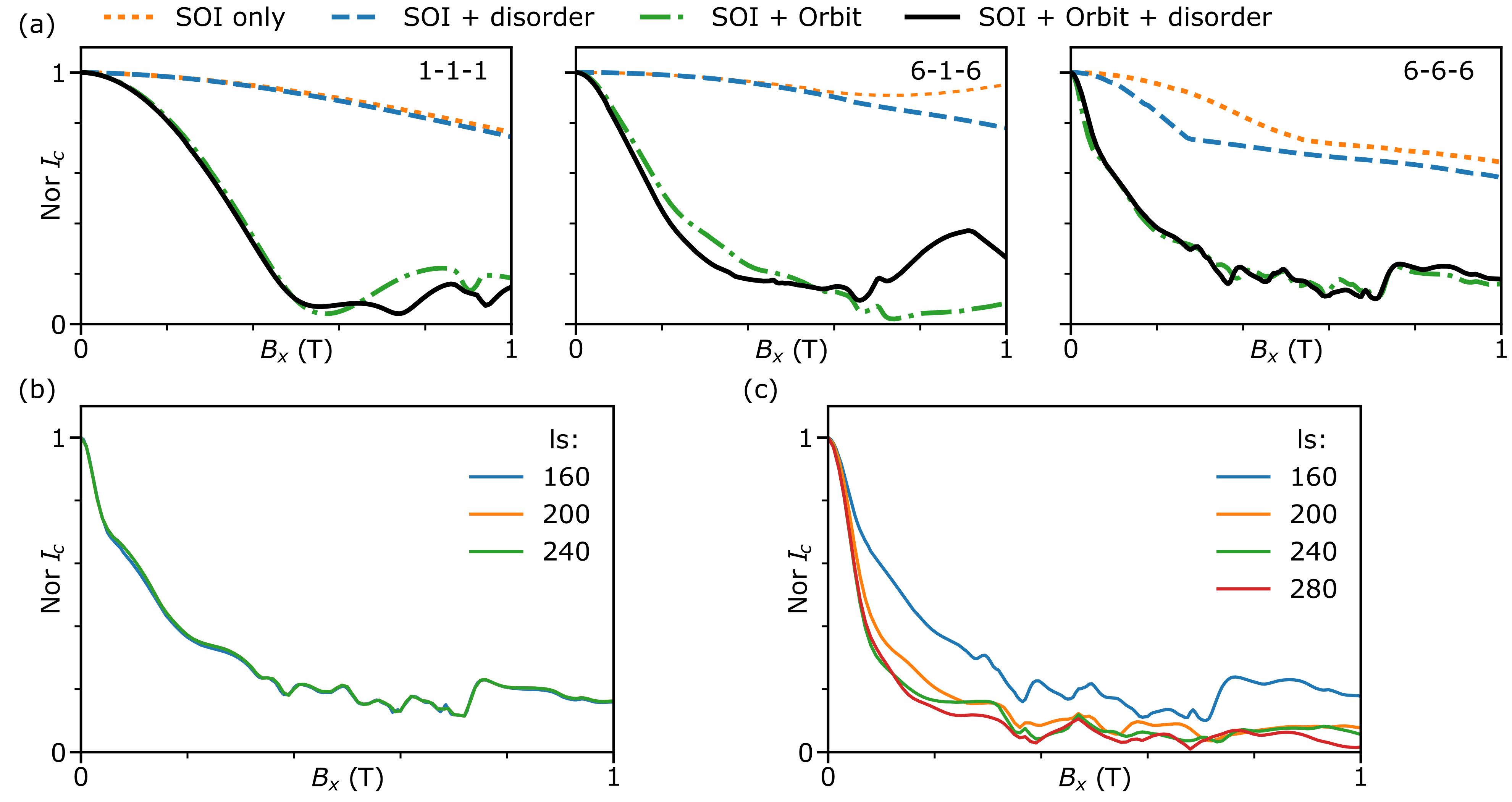}
\caption{\label{figs_sim_ic}. Parameters dependence of switching current diffraction pattern in the simulation. (a) Normalized critical current as a function of parallel field $B_x$ calculated with different parameter combination, the parameter space are same as Fig.~\ref{Fig5_Ic_sim} in the main text. The available modes in the lead/QPC/lead are labeled with number in each plot. (b) Junction length effect when there is no disorder distributed in the system ($\delta U$ = 0). (c) Junction length effect when there disorder is distributed in the system ($\delta U$ = 40~$\mathrm{meV}$). In (b) and (c), all three local chemical potential are set to be 20~$\mathrm{meV}$.}
\end{figure}

n Fig.~\ref{figs_sim_ic}, we study the impact of physical effects on supercurrent diffraction in the presence of an parallel field. Panel (a) displays the $I_c$ pattern when different parameter combinations are included in the simulation. The results that take into account spin-orbit interaction (SOI), orbital effects, and disorder are used to plot Fig.~\ref{Fig5_Ic_sim} in the main text, as these closely resemble real-world measurements.

In Fig.~\ref{figs_sim_ic}(b) and (c), we investigate how the length of the junction influences Orbital and Interference effects when the junction is adjusted to a fully open regime ($\mu_1, \mu_2, \mu_3$ = 20~$\mathrm{meV}$, all corresponding to six spin-full transverse modes). In panel (b), we did not introduce local disorder (setting $\delta U$ = 0) and found that the diffraction pattern of $I_c$ remains consistent for all junction lengths $l_s$. Thus, we conclude that the orbital effect is not influenced by the junction length. In panel (c), we introduced site disorder $\delta U$ = 40~$\mathrm{meV}$ and observed a faster decay of $I_c$ with larger junction lengths. This supports our argument regarding the distribution of disorder: electron transport in a longer junction is likely to encounter more scattering due to an increased number of defects introduced into the system, resulting in stronger suppression when a parallel field is applied.
\end{document}